\documentclass[12pt]{article}
\usepackage[iso-8859-1]{inputenx}
\usepackage{graphicx}
\usepackage[english]{babel}
\usepackage{amsmath,mathtools}
\usepackage{amssymb}
\usepackage[margin=1in]{geometry}
\usepackage{hyperref}
\usepackage{float}
\usepackage{multirow}
\usepackage{latexsym}
\usepackage{caption}
\usepackage{babelbib}
\usepackage{latexsym}
\usepackage[T1]{fontenc}
\usepackage{tabularx}
\usepackage{dcolumn}
\usepackage{eurosym}
\usepackage{setspace}
\usepackage{color}
\usepackage[usenames,dvipsnames]{xcolor}
\usepackage{geometry}
\usepackage{multicol}
\usepackage{caption}
\usepackage{subfigure}
\usepackage{amsthm}
\usepackage{stmaryrd}
\usepackage{enumitem}
\usepackage{physics}
\usepackage{bbold}
\usepackage[bb=ams]{mathalfa}
\usepackage{bbm}

\newcommand{\im}[2]{{#1}_{\scriptscriptstyle{#2}}}
\newcommand{\irm}[2]{{#1}_{\scriptscriptstyle{\text{#2}}}}

\newcommand{\te}[1]{\text{#1}}

\newcommand{\Lms}{\text{MS}}
\newcommand{\Lt}{\text{T}}

\newcommand{\Gv}{a}
\newcommand{\Gm}{\im{p}{\Gv}}
\newcommand{\Gvq}{\qs{\Gv}}
\newcommand{\Gmq}{\im{\qs{p}}{a}}
\newcommand{\KGv}{\phi}
\newcommand{\KGm}{\im{p}{\phi}}
\newcommand{\KGvq}{\qs{\phi}}
\newcommand{\KGmq}{\im{\qs{p}}{\phi}}
\newcommand{\MSv}{\nu}
\newcommand{\MSvq}{\qf{\nu}}
\newcommand{\MSmq}{\im{\qf{\pi}}{\MSv}}

\newcommand{\Gnv}{b}
\newcommand{\Gnm}{q'}
\newcommand{\KGnv}{\omega}
\newcommand{\KGnm}{u'}

\newcommand{\qpb}{(q,p)}
\newcommand{\qpbzero}{(\im{q}{0}, \im{p}{0})}

\newcommand{\M}{\mathcal{M}}

\newcommand{\Fs}[1]{\mathcal{F}_{s,#1}}

\newcommand{\Hs}[1]{\im{\mathcal{H}}{#1}}

\newcommand{\Hi}{\mathcal{H}}

\newcommand{\Hf}{\irm{\mathcal{H}}{p}}

\newcommand{\Ps}[1]{\im{\Gamma}{#1}}

\newcommand{\mMS}{\im{m}{\text{MS}}}
\newcommand{\mT}{\im{m}{\text{T}}}

\newcommand{\mMSt}{\im{\tilde{m}}{\text{MS}}}
\newcommand{\mTt}{\im{\tilde{m}}{\text{T}}}

\newcommand{\mPhi}{\im{m}{\Phi}}

\newcommand{\qs}[1]{\hat{#1}}
\newcommand{\qf}[1]{\pmb{#1}}

\newcommand{\h}[2]{\im{\qf{h}}{#1}^{#2}}

\newcommand{\heffscp}[1]{\im{\qf{h}}{\text{eff}#1}^{\qpb}}
\newcommand{\heff}[2]{\im{\qf{h}}{\text{eff}#1}^{#2}}

\newcommand{\aMS}[1]{\im{\qf{a}}{#1}}
\newcommand{\adMS}[1]{\im{\qf{a}}{#1}^{\dagger}}
\newcommand{\bT}[1]{\im{\qf{b}}{#1}}
\newcommand{\bdT}[1]{\im{\qf{b}}{#1}^{\dagger}}

\newcommand{\aMSt}[1]{\im{\tilde{\qf{a}}}{#1}}
\newcommand{\adMSt}[1]{\im{\tilde{\qf{a}}}{#1}^{\dagger}}
\newcommand{\bTt}[1]{\im{\tilde{\qf{b}}}{#1}}
\newcommand{\bdTt}[1]{\im{\tilde{\qf{b}}}{#1}^{\dagger}}

\newcommand{\Us}[1]{\im{\qs{1}}{#1}}
\newcommand{\Uf}[1]{\im{\qf{1}}{#1}}

\newcommand{\ef}[1]{\im{e}{#1}^{\qpb}}
\newcommand{\eft}[1]{\im{e}{#1}^{(\tilde{q},\tilde{p})}}
\newcommand{\e}[1]{\im{e}{#1}}
\newcommand{\efzero}[1]{\im{e}{#1}^{\qpbzero}}
\newcommand{\eftzero}[1]{\im{e}{#1}^{(\im{\tilde{q}}{0},\im{\tilde{p}}{0})}}

\newcommand{\eMSs}[1]{\im{e}{\lbrace...,\nMS{m}#1,...\rbrace}^{\scriptscriptstyle{\Lms}}}

\newcommand{\eTs}[1]{\im{e}{\lbrace...,\nT{M}#1,...\rbrace}^{\scriptscriptstyle{\Lt,\epsilon}}}

\newcommand{\Epd}[1]{E_{#1}^{\qpb}}

\newcommand{\pipd}[1]{\im{\qf{\pi}}{#1}^{\qpb}}

\newcommand{\pli}[1]{\im{\pmb{\pi}}{#1}}

\newcommand{\us}[2]{\im{\qf{u}}{#1}^{#2}}

\newcommand{\pp}{\varepsilon}
\newcommand{\scpr}[2]{\langle #1,#2 \rangle}
\newcommand{\spr}{\,\im{\star}{\pp}\,}
\newcommand{\sprT}{\,\im{\tilde{\star}}{\pp,\te{T}}\,}
\newcommand{\dr}{\mathrm{d}}

\newcommand{\aone}[2]{\alpha_{1,#1}^{\scriptscriptstyle{#2}}}
\newcommand{\atwo}[2]{\alpha_{2,#1}^{\scriptscriptstyle{#2}}}

\newcommand{\aonet}[2]{\tilde{\alpha}_{1,#1}^{\scriptscriptstyle{#2}}}
\newcommand{\atwot}[2]{\tilde{\alpha}_{2,#1}^{\scriptscriptstyle{#2}}}

\newcommand{\nd}{n_{\scriptscriptstyle{d}}}
\newcommand{\ndp}{\im{n}{d'}}
\newcommand{\nD}{n_{\scriptscriptstyle{\delta}}}
\newcommand{\nMS}[1]{\im{n}{\text{MS},\vec{#1},d}}
\newcommand{\nT}[1]{\im{n}{\text{T},\vec{#1},d'}}

\let\OLDthebibliography\thebibliography
\renewcommand\thebibliography[1]{
  \OLDthebibliography{#1}
  \setlength{\parskip}{1.5pt}
  \setlength{\itemsep}{2pt plus 2ex}
}

\title{{\rm Quantum Cosmological Backreactions IV:}\\
 {\rm Constrained Quantum Cosmological Perturbation Theory}} 
\author{
{\rm S. Schander}$^1$\thanks{{\rm 
susanne.schander@gravity.fau.de}},
{\rm T. Thiemann}$^1$\thanks{{\rm 
thomas.thiemann@gravity.fau.de}}\\
\\
{\rm $^1$ Inst. for Quantum Gravity, FAU Erlangen -- N\"urnberg,}\\
{\rm Staudtstr. 7, 91058 Erlangen, Germany}\\
}
\date{{\small\rm \today}}

\begin{document}

\maketitle

{\rm

\begin{abstract}
This is the fourth paper in a series of four in which we use space adiabatic methods in order to incorporate backreactions
among the homogeneous and between the homogeneous and inhomogeneous degrees of freedom in quantum cosmological perturbation theory. In this paper, we finally consider the gauge invariant scalar (Mukhanov-Sasaki) and tensor (primordial
gravitational wave) inhomogeneous perturbations of General Relativity coupled to an inflaton field which arise from a careful constraint analysis of this system up to second order in the perturbations. The simultaneous quantisation of the homogeneous and inhomogeneous degrees of freedom suggests the space adiabatic perturbation theory as an approximation scheme in order to capture the backreaction 
effects between these two sets of degrees of freedom. We are confronted with all the challenges at once that we found in the simpler models treated in this series of papers. We are able to compute these effects up to second order in the adiabatic parameter and find significant modifications as compared to earlier derivations of the effective quantum dynamics of the homogeneous sector.  

\end{abstract}

\newpage

\tableofcontents

\newpage

\section{Introduction}

The present paper is the culmination of this series of four papers, %\cite{1,2,3} 
\cite{Schander Thiemann I, Neuser Schander Thiemann II, Schander Thiemann III} in which we use 
space adiabatic methods, 
%\cite{4} 
\cite{Panati Spohn Teufel, Teufel}, together with the hybrid approach to quantum cosmology, \cite{5}, in order to capture quantum backreaction effects between the homogeneous and 
inhomogeneous degrees of freedom of both matter and geometry. 

Our motivation is as follows: Given the little experience that one has with the qualitative and quantitative aspects of potential quantum gravity effects for instance in quantum cosmology, we consider it as crucial to treat these tiny fingerprints of quantum gravity as carefully as possible. Present models for just the homogeneous sector such as Loop Quantum Cosmology
\cite{6} suggest a drastic departure from the cosmological big bang scenario in that the big bang is replaced by a big bounce
which should have phenomenological consequences. The test whether this prediction survives the incorporation of the quantum fluctuations of the inhomogeneous sector is obviously of outmost importance. In order to approach this question, one can 
consider the quantisation of classical cosmological perturbation theory, \cite{Mukhanov Article, Mukhanov Book}. This is precisely the idea of the hybrid approach \cite{5} which considers the inhomogeneous sector as a quantum field theory and the homogeneous sector as a quantum mechanical system which are in interaction. 

To precisely capture this interaction and to adequately estimate how it influences the effective dynamics of the homogeneous degrees of freedom, and in particular the fate of the classical big bang singularity is quite complicated. This task has been 
addressed using approximations \cite{8a,8b,8c} which rest on various assumptions. In this series we therefore have advertised the space adiabatic scheme as an unbiased approach to this question. It is ideally suited to the problem under investigation in that it lifts the idea of classical cosmological perturbation theory to the quantum level: The inhomogeneous fluctuations have only little effect on the homogeneous background which is the spatial average of the total field and thus acts like a ``centre of mass'' degree of freedom which moves much ``slower'' than the perturbation degrees of freedom, hence a space adiabatic perturbation theory scheme makes sense from the physical point of view. The adiabatic parameter unsurprisingly 
is related to ratio of mass scales of matter and geometry (Planck mass) and thus is expected to be rather tiny which makes the adiabatic expansion converge rather quickly.

The previous papers of this series have prepared us to deal with the many subtleties that one encounters when transferring the space adiabatic perturbation theory scheme developed for quantum systems with finitely many degrees of freedom to the field theory context, including:\\ 
1. The dependence of the Fock space vacua on the homogeneous degrees of freedom raises  
the question whether these are in fact members of the same Hilbert space, a prerequisite for the applicability of space adiabatic perturbation theory, \cite{9}. Fortunately, since exactly the same question also arises in the hybrid scheme \cite{Gomar Martin-Benito Mena Marugan, Gomar Cortez 
Martin-de Blas Mena Marugan}, this question can be answered in the affirmative once a suitable canonical transformation, exact up to second order in classical cosmological theory, has been carried out.\\
2. Since the Fock space vacua depend on both momentum and configuration variables of the homogeneous sector, the Born-Oppenheimer scheme is not applicable any longer
since the latter can only deal with a dependence on a commuting set of variables. Fortunately, space adiabatic perturbation theory is general enough to be able to treat this more general situation using Weyl quantisation techniques.\\
3. The Mukhanov-Sasaki mass and tensor mass terms are not positive in all regions of the homogeneous phase space which can lead to tachyonic modes. In our companion papers we have advertised several solutions to this problem (see also \cite{11})
including an ad hoc removal of the tachyonic modes or an ad hoc restriction of the classical phase space to regions where the mass terms are manifestly positive. In lack of a better proposal we adopt the second solution in the current paper for simplicity, however, this very interesting and important problem deserves further investigation in the future.\\
4. The mass terms and various other terms in the remaining contributions to the Hamiltonian constraint contain inverse powers of both momentum and configuration variables of the homogeneous sector. This is a new, more singular situation 
than one is used to even in Loop Quantum Cosmology and raises important domain questions for the resulting effective Hamiltonian for each Fock energy band. These questions are addressed and partly solved in a more general context in \cite{12} which we adopt 
for the current paper.\\
5. The backreaction contributions to that effective Hamiltonian involves a sum over inhomogeneous modes which can easily diverge. Fortunately, this does not occur up to second 
adiabatic order considered here. \\
\\
In this paper we use space adiabatic perturbation theory to compute the backreaction corrections to the effective Hamiltonian up to second order in the adiabatic expansion for each Fock energy band of the inhomogeneous sector.\\
The architecture of the present paper is as follows:\\

In the second, section we review the model, in particular following \cite{Gomar Martin-Benito Mena Marugan}.
In the third, we carry out the space adiabatic scheme.
In the fourth, we summarise and conclude.
In the appendix, we present some details of the straightforward but tedious calculations that lead to the effective Hamiltonian.

\section{Introduction of the Model}

In this work, we apply space adiabatic perturbation theory to standard cosmological perturbation theory with a gravitational metric field $g$, a massive real scalar field $\Phi$ as the matter content and a cosmological constant, $\Lambda \in \mathbb{R}^+$. After the split of the relevant degrees of freedom into a homogeneous and isotropic part and perturbations thereon, the aim will be to incorporate backreactions from the perturbative degrees of freedom onto the homogeneous and isotropic background degrees of freedom. \\
The model rests on a four-dimensional space time manifold $\M$, which we assume to be globally hyperbolic. The gravitational field $g$ on $\M$ is a two-times covariant, symmetric and non-degenerate tensor field with signature $(-,+,+,+)$ such that space time $(\M,g)$ is a Lorentzian manifold. According to a theorem by Geroch \cite{Geroch}, a globally hyperbolic manifold necessarily has the topology $\M \cong \mathbb{R} \times \sigma$, where $\sigma$ is a fixed three-dimensional manifold of arbitrary topology. In this work, we choose $\sigma$ to be the compact, flat three-torus $\mathbb{T}^3$ with side-length $L$. The purpose of this is to avoid technical problems in the ultraviolet regime of the field theory later on, and to model a flat universe which agrees with observations as long as the side-lengths of the torus are large compared to the size of the observable Universe.\\

The space adiabatic theory as developed in \cite{Panati Spohn Teufel,Teufel} requires a Hamiltonian formulation of the model, such that we adopt a $(3+1)$-split of space time as developed by Arnowitt, Deser and Misner, \cite{ADM}. Also \cite{Thomas} gives a detailed treatment of this approach. Due to the global hyperbolicity, $\M$ foliates into Cauchy surfaces, $\Sigma_t$, which are parametrized by a global time function, $t$. Let $n^{\mu}$ be the unit normal vector field to these hypersurfaces. Furthermore, let $N$ and $N_{\mu}$ be the (standard) lapse and shift function which parametrize the normal and the tangential part of the foliation of the hypersurfaces. Then, it is possible to formulate an initial value problem for the theory, and the task of specifying constraints or equations of motion for the metric field $g$, translates into finding the time evolution of the spatial metric $q_{\mu \nu} = g_{\mu \nu} + n_{\mu} n_{\nu}$ on the hypersurfaces induced by $g$. The complete definition of the initial value problem requires, in addition, the specification of the extrinsic curvature, $K_{\mu \nu} = q_{\mu}^{\rho} q_{\nu}^{\lambda} \nabla_{\rho} n_{\lambda}$, which is associated to the ``time derivative'' of $q$. $\nabla$ is the unique, torsion-free covariant derivative associated to the metric $g$. After pulling-back the tensor fields to $\mathbb{R}\times \mathbb{T}^3$ and denoting spatial indices on the spatial hypersurfaces with latin symbols, $a,b,c,..\in \left\lbrace1,2,3 \right\rbrace$, the Lagrange density at the bottom of our model is expressed by the sum of the Einstein-Hilbert Lagrange density $\irm{\mathcal{L}}{EH}$ of gravity and the scalar field Lagrange density $\im{\mathcal{L}}{\Phi},  $i.e., $\mathcal{L} = \irm{\mathcal{L}}{EH} + \im{\mathcal{L}}{\Phi}$, with,
\begin{equation} \label{eq:LEH}
\irm{\mathcal{L}}{EH} = \frac{1}{2 \kappa} \sqrt{|q|} N \left( R^{(3)} + K_{ab} K^{ab} - (K_a^a)^2 - 2 \Lambda \right),
\end{equation}
\begin{equation} \label{eq:LPhi}
\im{\mathcal{L}}{\Phi} =  \frac{1}{2\lambda} \sqrt{|q|} N \left( - \frac{1}{N^2} \dot{\Phi}^2 + 2 \frac{N^a}{N^2} \dot{\Phi} \partial_a \Phi + \left( q^{ab} - \frac{N^a N^b}{N^2} \right) \partial_a \Phi \partial_b \Phi + \im{m}{\Phi}^2 \Phi^2 \right).
\end{equation}
Here, $\kappa = 8 \pi G$ is the gravitational coupling constant, $\lambda$ is the coupling constant of the scalar field, $\im{m}{\Phi}$ is the mass parameter of the scalar field, and $R^{(3)}$ is the curvature scalar associated with the three-metric $q$ and its Levi-Civita covariant derivative, $D$. \\

The cosmological setting in this work, suggests to consider a homogeneous and isotropic restriction of the gravitational theory at hand, up to small deviations. These symmetry reductions imply that the only remaining degrees of freedom for the homogeneous and isotropic part of the system are the zeroth order lapse function $\im{N}{0}(t)$ and the scale factor $a(t)$, associated with the zeroth order spatial metric $\prescript{\scriptscriptstyle{0\!}}{}{q}(x^i) = a^2(t)\,\prescript{\scriptscriptstyle{0\!}}{}{\tilde{q}}(x^i)$, where we introduced the fixed spatial metric $\prescript{\scriptscriptstyle{0\!}}{}{\tilde{q}}(x^i)$ on the spatial hypersurfaces. A Hamiltonian analysis shows that the lapse function is a Lagrange multiplier of the system and has, hence, no dynamical features. This affirms the arbitrariness of the hypersurface foliation. The next step consists in introducing perturbations of the homogeneous and isotropic metric tensor, and for the scalar field. In this respect, it is convenient to decompose the perturbative fields into scalar, vector and tensor parts according to their properties regarding $SO(3)$-transformations. This is reasonable since the respective equations of motion decouple. Note that the procedure of introducing perturbative fields on a homogeneous and isotropic background introduces a gauge freedom for the perturbations since the choice of coordinates is a priori arbitrary. A detailed analysis of cosmological perturbation theory within the Hamiltonian framework for closed FRW universes can be found in \cite{Halliwell Hawking}. There, however, the gauge freedom of the perturbations was fixed by choosing one particular gauge. Regarding the scalar part of the perturbations, we refer to \cite{Gomar Martin-Benito Mena Marugan} where the authors use gauge-invariant Mukhanov-Sasaki perturbations. For the tensor perturbations, we point to \cite{Martinez Olmedo}. Similar to the definitions in \cite{Gomar Martin-Benito Mena Marugan,Martinez Olmedo}, we define the perturbed lapse, shift, spatial metric and matter scalar field respectively as,
\begin{align}
N(t,x^i) &= \im{N}{0}(t) + a^3(t)\, g(t,x^i) \label{eq:Lapse} \\
N_a(t,x^i) &=: a^2(t)\,D_a\, k(t,x^i) + a^2(t)\,\epsilon_a^{\; b c} D_b\,k_c(t,x^i) \label{eq:Shift} \\
q_{ab}(t,x^i) &=:  a^2(t) \left[\left(1+2\,\alpha(t,x^i) \right) \!\prescript{\scriptscriptstyle{0}\!}{}{\tilde{q}}_{ab}(x^i) + 6 \left( D_a D_b - \frac{1}{3} \prescript{\scriptscriptstyle{0\!}}{}{\tilde{q}}_{ab}(x^i) D_c D^c \right) \beta(t, x^i) \right. \nonumber \\
& ~~~~~~~~~~~~~ \left.+\, 2 \sqrt{6} \, T_{ab}(t, x^i) + 4 \sqrt{3} D_{(a} V_{b)}(t,x^i) \right],\label{eq:Perturbation Metric} \\
\Phi(t,x^i) &=: \phi(t) + f(t,x^i). \label{eq:phi}
\end{align}
Recall that the variables $(\im{N}{0},a,\phi)$ refer to the homogeneous and isotropic part of the system. Concerning the perturbations, we denote the scalar fields by $(g, k, \alpha, \beta, f )$, the vector degrees of freedom by $v_a$ and $k_a$, and the tensor field perturbations by $t_{a b}$. For notational reasons, we introduce $\tilde{k}:=\Delta k$ and $\tilde{k}_a:= \epsilon_a^{\;bc} D_b k_c$ as new degrees of freedom associated with the shift. In contrast to the proceeding in \cite{Gomar Martin-Benito Mena Marugan,Martinez Olmedo}, we stick to a space time representation of the perturbative fields instead of choosing a particular Fourier mode decomposition.\\

The next step consists in inserting the perturbed variables of the definitions (\ref{eq:Lapse},\ref{eq:Shift},\ref{eq:Perturbation Metric},\ref{eq:phi}) into the Lagrange density (\ref{eq:LEH},\ref{eq:LPhi}), and then to expand the Lagrangian and the action functional $S$ up to second order in the perturbations. Because the three-torus does not have a boundary, total divergences vanish in the computations. The resulting action does neither depend on the velocities of the lapse variables $\im{N}{0}$, $g$, nor on the velocities of the shift variables $\tilde{k}$, $\tilde{k}_a$. This implies that lapse and shift are Lagrange multipliers and will hence be associated to primary constraint equations in the Hamiltonian picture.  In order to pass over to the Hamiltonian picture, we perform a Legendre transformation in the lines of \cite{Halliwell Hawking,Gomar Martin-Benito Mena Marugan}. Thereby, we define the conjugate momenta $(\im{p}{a}, \im{p}{\phi})$ for the homogeneous and isotropic degrees of freedom $(a,\phi)$, as well as the conjugate momenta $(\im{\pi}{\alpha}, \im{\pi}{\beta}, \im{\pi}{f}, \pi^a_{\scriptscriptstyle{v}}, \pi^{ab}_{\scriptscriptstyle{t}})$ assigned to the perturbation fields $(\alpha, \beta, f, v_a, t_{ab})$. The variables $\im{N}{0}$, $g$, $\tilde{k}$ and $\tilde{k}_a$ induce the lapse and shift primary constraints $\im{\Pi}{0}^{\im{N}{0}}$, $\im{\Pi}{1}^{g}$, $\im{\Pi}{1}^{\tilde{k}}$ and $\im{\Pi}{1}^{\tilde{k}_a,b}$. The Legendre transformation yields a Hamiltonian density of the form,
\begin{align}
\mathcal{H} =&~ \im{N}{0} \left[ \im{\mathcal{H}}{0} + \im{\mathcal{H}}{2}^{\text{s}} + \im{\mathcal{H}}{2}^{v} + \im{\mathcal{H}}{2}^{t} \right] + g \cdot \im{\mathcal{H}}{1}^{g} + \tilde{k}_a \cdot \im{\mathcal{H}}{1}^{\tilde{k}_d,a} + \tilde{k} \cdot \im{\mathcal{H}}{1}^{\tilde{k}} \nonumber \\
& + \im{\lambda}{\im{N}{0}}\cdot \im{\Pi}{0}^{\im{N}{0}} + \im{\lambda}{g} \cdot \im{\Pi}{1}^{g} + \im{\lambda}{\tilde{k}}\cdot\im{\Pi}{1}^{\tilde{k}} + \im{\lambda}{\tilde{k}_a,b} \cdot \im{\Pi}{1}^{\tilde{k}_a\!,b} \label{eq:H original}
\end{align}
Here, $\im{\mathcal{H}}{0}$ denotes the zeroth order Hamiltonian contribution associated with the completely homogeneous and isotropic model. The contributions $\im{\mathcal{H}}{2}^{\text{s}}$, $\im{\mathcal{H}}{2}^{v}$ and $\im{\mathcal{H}}{2}^{t}$ are of second order in the perturbations and contain only scalar, vector and tensor variables respectively. The terms $\im{\mathcal{H}}{1}^g$, $\im{\mathcal{H}}{1}^{\tilde{k}_d,a}$ and $\im{\mathcal{H}}{1}^{\tilde{k}}$ represent first order contributions which factorize with the respective lapse and shift variables. The second line only lists the primary constraints associated with lapse and shift and their Lagrange multipliers $\im{\lambda}{\im{N}{0}}$, $\im{\lambda}{g}$, $\im{\lambda}{\tilde{k}}$ and  $\im{\lambda}{\tilde{k}_a,b}$. \\

In a next step, the aim is to perform a Dirac analysis in order to derive the dynamical properties of the system. Thereby, we encounter several difficulties: First, the perturbation variables that we introduced are not all gauge-invariant. Therefore, a canonical transformation to gauge-invariant variables would be necessary in order to have a covariant theory of the perturbations. Indeed, it is straightforward to introduce the gauge-invariant Mukhanov-Sasaki variable $\nu$ in the scalar sector of the perturbations, see for example \cite{Mukhanov Article,Mukhanov Book}. Thereby, however, we perform a transformation for the perturbations only and in order to preserve the canonical structure of the system, it is mandatory to find a suitable transformation for the homogeneous and isotropic variables, too. This appears to be a cumbersome mission. However, Gomar, Mart\'{i}n-Benito and Mena Marug\'{a}n have shown in \cite{Gomar Martin-Benito Mena Marugan} that it is possible to find a transformation for the homogeneous and isotropic degrees of freedom which preserves the canonical structure of the system up to second order in the perturbations. The same has been done by Mart\'{i}nez and Olmedo in \cite{Martinez Olmedo} for the tensor degrees of freedom. We employ these transformations in this work. \\
The second difficulty regarding the Dirac algorithm concerns the closure of the constraint algebra. In general, the algorithm might entail a large number of constraints which are not well manageable. The idea, put forward in \cite{Gomar Martin-Benito Mena Marugan,Martinez Olmedo} which we will also apply in this work, is to use some of the secondary constraints of the Dirac algorithm as the canonical variables themselves. Thereby, the Dirac algorithm becomes partly trivial just by implementing the first set of secondary constraints. This will be demonstrated in the sequel. \\
In summary, the aim of the following procedure is then threefold; First, we wish to introduce gauge-invariant variables for the perturbations in order to circumvent problems occurring for coordinate changes. Second, we aim at keeping the canonical structure of the theory, at least up to second order in the cosmological perturbations. For the latter purpose, we will review the Dirac algorithm for constrained systems and implement additional transformations for the homogeneous and isotropic degrees of freedom. In particular, we modify the homogeneous variables by adding second order contributions of the perturbations.
Third, we wish to construct a theory whose dynamics will be unitarily implementable at the quantum level. Therefore, we consider further canonical transformations with respect to the perturbations. Their effects on the homogeneous variables will be taken into account accordingly. Following \cite{Gomar Martin-Benito Mena Marugan,Martinez Olmedo}, the formalism proceeds as follows:\\

As a starting point, we consider the homogeneous and isotropic degrees of freedom as to be non-dynamical background variables. This offers the possibility to introduce perturbation variables which build a canonical set with respect to the dynamical, perturbative system only. \\
We start with the canonical pair of the tensor perturbations $(t_{ab}, \im{\pi}{t}^{ab})$ which is already gauge-invariant. However, we aim at obtaining classical perturbation variables whose dynamics is unitarily implementable in the quantum realm. As shown for example in \cite{Gomar Cortez Martin-de Blas Mena Marugan,FernandezMendez:2012sr,Gomar:2012xn,Cortez Mena Marugan Velhinho}, this simply amounts to eliminating contributions in the Hamiltonian which couple the perturbation variables with their respective momenta. In this way, the final Hamiltonian at second order will only consist of terms proportional to squares of the perturbation variables or squares of the perturbation momenta after a suitable transformation. In other words, the Hamiltonian has the form of a sum of harmonic oscillators with masses and frequencies that possibly depend on the homogeneous and isotropic degrees of freedom. Indeed, these transformations guarantee the unitarity of the perturbations quantum dynamics when considered in a semiclassical framework of a quantum field theory on a curved space time. We employ the transformations found by Martin\'{e}z and Olmedo in \cite{Martinez Olmedo} for the tensor perturbations and transform the homogeneous degrees of freedom accordingly by adding second order contributions. These yield additional terms of second order tensor perturbations in the Hamiltonian which will be absorbed into $\im{\mathcal{H}}{2}^t$. Accordingly, we denote the new tensor Hamiltonian as $\im{\tilde{\mathcal{H}}}{2}^t$. Furthermore, the transformations result into a shift of the lapse function by second order contributions which will be taken into account by a function denoted as $\im{\tilde{N}}{2}$. \\
Regarding the vector perturbations, we can identify the constraints $\im{\mathcal{H}}{1}^{k_d\!,a}$ and their conjugate variables $C_{\scriptscriptstyle{1},a}^{k_d} = 2 \sqrt{3}\,v_a$ as canonical pairs. The transformation for these perturbation variables entails a transformation for the homogeneous degrees of freedom in order to keep the (almost) canonical structure. Both transformations result in a new second order vectorial part of the Hamiltonian, $\im{\tilde{\mathcal{H}}}{2}^v$, which is proportional to the constraint $\im{\mathcal{H}}{1}^{k_d\!,a}$ itself. Thus, if we demand that $\im{\mathcal{H}}{1}^{k_d\!,a}$ vanishes as a constraint, this implies that $\im{\tilde{\mathcal{H}}}{2}^v$ vanishes, too.  \\

In the scalar sector, we employ the Mukhanov-Sasaki scalar field $\nu$. As suggested in \cite{Gomar Martin-Benito Mena Marugan}, it is helpful to additionally consider the first order constraints $\im{\mathcal{H}}{1}^g$ and $\im{\mathcal{H}}{1}^{\tilde{k}}$ as new perturbation variables. Since these constraints do not commute with respect to the perturbation Poisson brackets, we shift $\im{\mathcal{H}}{1}^g$ by a linear term in the perturbations and we obtain the final variable $\im{\tilde{\mathcal{H}}}{1}^g$, which commutes with $\im{\mathcal{H}}{1}^{\tilde{k}}$, if we only consider the perturbations as dynamical degrees of freedom. This procedure yields another shifting of the lapse function $\im{\tilde{N}}{2}$. In a next step, we construct the conjugate variables with respect to the inhomogeneous Poisson brackets only, denoting them by $\im{\pi}{\nu}$, $\im{C}{1}^g$ and $\im{C}{1}^{\tilde{k}}$. The new canonical pairs in the scalar sector of the perturbations are thus, $(\nu, \im{\pi}{\nu})$, $(\im{C}{1}^g, \im{\tilde{\mathcal{H}}}{1}^g)$ and $(\im{C}{1}^k, \im{\mathcal{H}}{1}^{\tilde{k}})$. Finally, we complete the transformation in the homogeneous sector by adding second order contributions to the initial canonical pairs. The implementation of the transformations yield new contributions to $\mathcal{H}$: some of them include only the Mukhanov-Sasaki canonical variables and we correspondingly absorb them into a new second order scalar Hamiltonian $\im{\tilde{\mathcal{H}}}{2}^s$; another contribution is proportional to the zeroth order Hamiltonian $\im{\mathcal{H}}{0}$ such that it is possible to absorb them into $\im{\tilde{N}}{2}$. In addition, the transformations result into new second order contributions which are proportional to the linear constraints $\im{\tilde{\mathcal{H}}}{1}^g$ and $\im{\mathcal{H}}{1}^{\tilde{k}}$. We denote these contributions as $\im{G}{1}$ and $\im{K}{1}$ respectively.   \\

The transformations yield in total the following Hamiltonian density,
\begin{align}
\mathcal{H} =& \left(\im{N}{0} + \im{\tilde{N}}{2}\right) \cdot \left[ \im{\mathcal{H}}{0} + \im{\tilde{\mathcal{H}}}{2}^s + \im{\tilde{\mathcal{H}}}{2}^v + \im{\tilde{\mathcal{H}}}{2}^t\right] + \left(g+\im{G}{1} \right)\! \cdot \! \im{\tilde{\mathcal{H}}}{1}^g + \left(\tilde{k}+\im{K}{1} \right)\! \cdot \!\im{\mathcal{H}}{1}^{\tilde{k}} + \tilde{k}_a \cdot \im{\mathcal{H}}{1}^{\tilde{k}_d,a} \nonumber \\
&~~ + \im{\lambda}{\im{N}{0}} \!\!\cdot\!  \im{\Pi}{0}^{\im{N}{0}} + \im{\lambda}{g}\!\!\cdot\! \im{\Pi}{1}^g + \im{\lambda}{\tilde{k}}\!\!\cdot\! \im{\Pi}{1}^{\tilde{k}} + \lambda_{\scriptscriptstyle{\tilde{k}_b},a}\! \! \cdot\! \im{\Pi}{1}^{\tilde{k}_b,a} . \label{eq:H final}
\end{align}
The second line of formula \eqref{eq:H final} accounts for the primary constraints $\mathbf{\Pi} :=(\im{\Pi}{0}^{\im{N}{0}}, \im{\Pi}{1}^g, \im{\Pi}{1}^{\tilde{k}}, \im{\Pi}{1}^{\tilde{k}_b,a})$ with their respective Lagrange multipliers, $(\im{\lambda}{\im{N}{0}}, \im{\lambda}{g}, \im{\lambda}{\tilde{k}}, \lambda_{\tilde{k}_b,a})$. These primary constraints already appeared in \eqref{eq:H original} and have remained unchanged under the preceding transformations. The system restricts to the submanifold of phase space defined by the primary constraints,
\begin{equation}
\im{\Pi}{0}^{\im{N}{0}} = 0,~ \im{\Pi}{1}^{g} = 0,~\im{\Pi}{1}^{\tilde{k}} = 0,~ \im{\Pi}{1}^{\tilde{k}_b,a} = 0. \label{eq:Primary Constraints}
\end{equation}
Subsequently, the associated Lagrange multipliers can be chosen arbitrarily. In a second step, consistency of the dynamics requires that the primary constraints remain zero under the evolution generated by the full Hamiltonian $\mathcal{H}$. This requirement gives rise to the secondary constraints,
\begin{align}
\left\lbrace \mathcal{H}, \im{\Pi}{0}^{\im{N}{0}} \right\rbrace &= \im{\mathcal{H}}{0} + \im{\tilde{\mathcal{H}}}{2}^s +\im{\tilde{\mathcal{H}}}{2}^v + \im{\tilde{\mathcal{H}}}{2}^t\approx 0, \label{eq:Sec Constraint 1} \\
\left\lbrace \mathcal{H}, \im{\Pi}{1}^g \right\rbrace &= \im{\tilde{\mathcal{H}}}{1}^g \approx 0, \label{eq:Sec Constraint 2}\\
\left\lbrace \mathcal{H}, \im{\Pi}{1}^{\tilde{k}} \right\rbrace &= \im{\mathcal{H}}{1}^{\tilde{k}} \approx 0. \label{eq:Sec Constraint 3a}\\
\left\lbrace \mathcal{H}, \im{\Pi}{1}^{\tilde{k}_b,a} \right\rbrace &= \im{\mathcal{H}}{1}^{\tilde{k}_b,a} \approx 0. \label{eq:Sec Constraint 3b}
\end{align}
Note that now, the Poisson brackets include the dynamics with respect to \emph{all} canonical pairs of the transformed system, both the homogeneous and the inhomogeneous ones. Indeed, the formalism allows us to compute the dynamics for the full system in the standard Hamiltonian framework. \\
The next step consists in checking whether the secondary constraints in \eqref{eq:Sec Constraint 1} - \eqref{eq:Sec Constraint 3b} are preserved under the dynamics of $\mathcal{H}$, or if they entail further secondary constraints. The computations are trivial since the preceding transformations imply that the first order constraints $(\im{\tilde{\mathcal{H}}}{1}^g, \im{\mathcal{H}}{1}^{\tilde{k}},\im{\mathcal{H}}{1}^{\tilde{k}_d,a})$ are canonical variables, and hence commute with all other variables except for their conjugate variables, $(\im{C}{1}^g, \im{C}{1}^{\tilde{k}}, \im{C}{1}^{\tilde{k}_d,a})$. Indeed, $\im{C}{1}^g$ appears in $\mathcal{H}$ within the first order functions $\im{G}{1}$ and $\im{K}{1}$ and thus, entails non-vanishing Poisson brackets with $\im{\tilde{\mathcal{H}}}{1}^g$. Since these Poisson brackets enter however with a constraint factor, they vanish at least weakly, 
\begin{align}
\left\lbrace \mathcal{H},\im{\mathcal{H}}{0} + \im{\tilde{\mathcal{H}}}{2}^s + \im{\tilde{\mathcal{H}}}{2}^v + \im{\tilde{\mathcal{H}}}{2}^t\right\rbrace &= 0, \label{eq:Sec Constraint 4} \\
\left\lbrace \mathcal{H},  \im{\tilde{\mathcal{H}}}{1}^g \right\rbrace &= \left\lbrace \im{G}{1} , \im{\tilde{\mathcal{H}}}{1}^g  \right\rbrace \im{\tilde{\mathcal{H}}}{1}^g + \left\lbrace \im{K}{1} , \im{\tilde{\mathcal{H}}}{1}^g  \right\rbrace \im{\mathcal{H}}{1}^k \approx 0, \label{eq:Sec Constraint 5}\\
\left\lbrace \mathcal{H}, \im{\mathcal{H}}{1}^k\right\rbrace &= \left\lbrace \im{G}{1} , \im{\mathcal{H}}{1}^k  \right\rbrace \im{\tilde{\mathcal{H}}}{1}^g + \left\lbrace \im{K}{1} , \im{\mathcal{H}}{1}^k  \right\rbrace \im{\mathcal{H}}{1}^k = 0.  \label{eq:Sec Constraint 6}
\end{align}
In summary, the constraint algebra closes and we are able to solve the dynamics of the system. Therefore, the primary constraints \eqref{eq:Primary Constraints}, as well as the  secondary constraints,
\begin{align}
\im{\tilde{\mathcal{H}}}{1}^g = 0,~\im{\tilde{\mathcal{H}}}{1}^{\tilde{k}} = 0,~ \im{\mathcal{H}}{1}^{\tilde{k}_b,a} = 0,~\im{\mathcal{H}}{0} + \im{\tilde{\mathcal{H}}}{2}^s + \im{\tilde{\mathcal{H}}}{2}^v + \im{\tilde{\mathcal{H}}}{2}^t =0,
\end{align}
must be satisfied on the constraint surface. Since $\im{\tilde{\mathcal{H}}}{1}^g$, $\im{\tilde{\mathcal{H}}}{1}^{\tilde{k}}$ and $\im{\mathcal{H}}{1}^{\tilde{k}_b,a}$ have simply become canonical momenta after the transformation such that we don't need to analyse these constraints further. We recall that the second order constraint $\im{\tilde{\mathcal{H}}}{2}^v$ is zero whenever, $\im{\mathcal{H}}{1}^{\tilde{k}_b,a} = 0$, holds. Hence, the only non-trivial constraint of the cosmological system amounts to be,
\begin{equation}
\im{\mathcal{H}}{0} + \im{\tilde{\mathcal{H}}}{2}^s +\im{\tilde{\mathcal{H}}}{2}^t =0. \label{eq:Hamilton Constraint}
\end{equation}
It is the object of interest in this work and we specify its contributions in the following. 

\section{The Hamilton Constraint}
First, we recall that the system variables consist, on the one hand, of the homogeneous and isotropic canonical pairs, $(\tilde{a},\im{\tilde{p}}{a})$ and $(\tilde{\phi}, \im{\tilde{p}}{\phi})$. These are associated with the standard cosmological scale factor $a$ and the homogeneous and isotropic part of the scalar matter field $\phi$ but have been shifted by second order contributions in the cosmological perturbations in order to maintain the (almost) canonical structure of the system. In what follows, we omit the dashes for simplicity. \\
In order to make space adiabatic perturbation theory work at the technical level, we rescale several variables. Therefore, we define the dimensionless parameter $\varepsilon$ by means of the ratio of the gravitational and the scalar matter coupling constant,
\begin{equation}
\varepsilon^2 := \frac{\kappa}{\lambda}.
\end{equation}
Since the gravitational coupling constant appears to be about thirty orders of magnitude smaller than the Standard Model coupling constants, $\varepsilon$ can be identified as a small, perturbative parameter for our space adiabatic perturbative analysis. The scheme suggests to rescale the homogeneous degrees of freedom according to, 
\begin{equation}
\im{\breve{p}}{a} := \pp^2\,\im{p}{a},~~ \im{\breve{p}}{\phi} :=  \pp\, \im{p}{\phi}, \label{eq:Scaling hom}
\end{equation}
as well as the Mukhanov-Sasaki field variables $(\nu, \im{\pi}{\nu})$ and the tensor field variables $(t_{ab}, \im{\pi}{t}^{ab})$ following,
\begin{equation}
\breve{\nu} := \frac{\nu}{\pp}, ~~ \im{\breve{\pi}}{\nu} := \pp\,\nu~~~~ \text{and}~~~~ \breve{t}_{ab} := \frac{t_{ab}}{\pp^2},~~ \im{\breve{\pi}}{t}^{ab} := \pp^2\, \im{\pi}{t}^{ab}. \label{eq:Scaling inhom}
\end{equation} 
We directly relabel the rescaled variables by removing the breves such that the notation remains as simple as possible. 
Note that the transformations for the perturbation fields, \eqref{eq:Scaling inhom}, are canonical, while the canonical structure of the homogeneous degrees of freedom change due to the rescaling in \eqref{eq:Scaling hom}. This becomes evident when considering the canonical quantum commutation relations in the following.\\

Space adiabatic perturbation theory considers the whole system as a quantum system; it does not rely on any semiclassical approximations. In the homogeneous sector of the model, we introduce hats for indicating quantum operators and we denote the Hilbert spaces of the gravitational subsystem as $\Hs{\Gv}$ and the matter subsystem as $\Hs{\KGv}$. Thus, the operators for the homogeneous sector are formally defined as $(\qs{a},\im{\qs{p}}{a}, \qs{\phi}, \im{\qs{p}}{\phi})$ and the commutation relations are given by,
\begin{equation}
\im{\left[ \Gvq, \Gmq \right]}{\Gv} = \frac{i\,\varepsilon^2}{L^3} \Us{\Hs{\Gv}}, ~~~~ \im{\left[ \KGvq, \KGmq \right]}{\KGv} = \frac{i\,\varepsilon}{L^3} \Us{\Hs{\KGv}}. 
\end{equation}
Note that the homogeneous degrees of freedom are considered as center of mass degrees of freedom on the spatial manifold $\mathbb{T}^3$, as it has been argued in \cite{Schander Thiemann I}. 
Indeed, the application of space adiabatic perturbation theory to our cosmological model considers the homogeneous and isotropic degrees of freedom as the ones whose canonical structure becomes rescaled by a very small parameter. It is thus reasonable to use \emph{phase space} quantum mechanics for treating these variables in the full quantum regime of the model. On the other hand, bold characters indicate quantum operators of the inhomogeneous system, in particular $(\qf{\nu}, \im{\qf{\pi}}{\nu}, \qf{t}_{ab}, \im{\qf{\pi}}{t}^{ab})$. We denote the Hilbert space of the Mukhanov-Sasaki quantum system as $\Hs{\Lms}$, the tensor Hilbert space as $\Hs{\Lt}$, and the perturbations Hilbert space arises as the tensor product of the two latter, $\Hs{\text{p}} = \Hs{\Lms} \otimes \Hs{\Lt}$. The canonical commutation relations for the perturbation fields are,
\begin{equation}
\im{\left[ \MSvq(t,\vec{x}), \MSmq(t,\vec{y}) \right]}{\Lms} = \delta^3(\vec{x},\vec{y}) \Uf{\Hs{\Lms}}, ~~~~  \im{\left[ \qf{t}_{ab}(t,\vec{x}), \im{\qf{\pi}}{t}^{cd}(t,\vec{y})\,\right]}{\Lt} = \delta_{(a}^c \delta_{b)}^d\,\delta^3(\vec{x},\vec{y}) \,\Uf{\Hs{\Lt}}.
\end{equation}
In the following, we explicitly employ quantum operators with respect to the perturbative part, while maintaining the phase space picture of the homogeneous part, thus treating the homogeneous variables at first as real-valued parameters. This leads us to introduce operator valued functions on the homogeneous phase space $\Ps{\!\!\text{hom}}$, also denoted as ``symbols'' and we generically write for the class of such functions, $S(\Ps{\!\!\text{hom}}, \mathcal{L}(\Hs{\text{p}}))$. With the scalings of the canonical variables in \eqref{eq:Scaling hom}, \eqref{eq:Scaling inhom}, and after a multiplication by the overall factor $\pp^2$, the quantum Hamilton constraint in \eqref{eq:Hamilton Constraint} is given in the lines of space adiabatic perturbation theory as a symbol function in $S(\Ps{\!\!\text{hom}}, \Hs{\text{p}})$ by, 
\begin{align} \label{eq:h}
\qf{h} := \pp^2 \left( \im{\qf{\mathcal{H}}}{0} + \im{\tilde{\qf{\mathcal{H}}}}{2}^s + \im{\tilde{\qf{\mathcal{H}}}}{2}^t \right) =&\, L^3 \left(- \frac{\lambda}{12} \frac{\Gm^2}{\Gv} + \lambda\, \frac{\KGm^2}{2\,\Gv^3} + \frac{1}{2\,\lambda}\,\varepsilon^2\,\mPhi^2 \Gv^3\,\KGv^2 + \frac{\Lambda}{\lambda}\,\Gv^3 \right)\cdot \Uf{\Hs{\text{p}}} \\
& + \frac{1}{2\,\Gv} \int_{\mathbb{T}^3} \dr^3x\,\left( \frac{\lambda\, \MSmq^2}{\sqrt{\prescript{\scriptscriptstyle{0\!}}{}{\tilde{q}}}}  + \MSvq \cdot \pp^4 \left( - \frac{\sqrt{\prescript{\scriptscriptstyle{0\!}}{}{\tilde{q}}}}{\lambda} \Delta + \mMS^2 \right) \MSvq \right) \otimes \Uf{\Hs{\Lt}} \nonumber\\
& + \frac{\Uf{\Hs{\Lms}}}{2\,\Gv} \otimes \int_{\mathbb{T}^3} \dr^3x\, \left( \frac{\lambda\,\im{\qf{\pi}}{t}^{ab} \qf{\pi}_{\scriptscriptstyle{t},ab}}{6\sqrt{\prescript{\scriptscriptstyle{0\!}}{}{\tilde{q}}}}  + \qf{t}^{ab} \pp^4 \left( - 3 \frac{\sqrt{\prescript{\scriptscriptstyle{0\!}}{}{\tilde{q}}}}{\lambda} \Delta + (\pp\,\mT)^2  \right) \qf{t}_{ab}\!\right)\!. \nonumber
\end{align}
Thereby, we introduced the masses associated with the Mukhanov-Sasaki and the tensor variables,
\begin{align}
\mMS^2 &:= \left(- \frac{\Gm^2}{18\,\Gv^2} + \frac{7\,\KGm^2}{2\,\Gv^4} -  12\,\pp\,\frac{\sqrt{\prescript{\scriptscriptstyle{0\!}}{}{\tilde{q}}}^{\,2}}{\lambda^2}\,\mPhi^2\,\frac{\Gv\,\KGv\,\KGm}{\Gm} - 18\,\frac{\KGm^4}{\Gv^6\,\Gm^2} + \frac{\sqrt{\prescript{\scriptscriptstyle{0\!}}{}{\tilde{q}}}^{\,2}\,\mPhi^2}{\lambda^2}\, \Gv^2 \right) \frac{\lambda}{\sqrt{\prescript{\scriptscriptstyle{0\!}}{}{\tilde{q}}}}\\
(\pp\,\mT)^2 &:= \left(\frac{\Gm^2}{6\,\Gv^2} - 3\,\pp^2\,\frac{\sqrt{\prescript{\scriptscriptstyle{0\!}}{}{\tilde{q}}}^{\,2}\,\mPhi^2}{\lambda^2}\,\Gv^2 \KGv^2 - 6 \frac{\sqrt{\prescript{\scriptscriptstyle{0\!}}{}{\tilde{q}}}^{\,2} \Lambda}{\lambda^2}\, \Gv^2 \right) \frac{\lambda}{\sqrt{\prescript{\scriptscriptstyle{0\!}}{}{\tilde{q}}}}.
\end{align}
Note that, in fact, the Hamiltonian symbol $\qf{h}$ comprises only an off-set energy contribution which depends parametrically on the homogeneous phase space variables as well as field contributions of the standard Klein-Gordon type for the Mukhanov-Sasaki and the tensor fields. These latter contributions have masses and frequencies which depend on the homogeneous variables, too. Since we assume the space manifold to be compact, a discrete mode decomposition for the fields is available. The modes take hence values in $\mathbbm{k}:=\mathbb{Z}^3\setminus \lbrace 0 \rbrace$, and in order to avoid repetition of modes, the first non-vanishing component is strictly positive. Also recall that the tensor field only carries two independent degrees of freedom, corresponding to the two polarizations of the tensor modes. These will be labeled by the index $\epsilon = \lbrace +,- \rbrace$. The state space associated with the Mukhanov-Sasaki and the tensor systems is the tensor product of the symmetric Fock spaces,
\begin{equation}
\Hs{\text{p}} = \mathcal{F}_{s,\scriptscriptstyle{\Lms}}(\ell^2(\mathbbm{k})) \bigotimes_{\epsilon =\lbrace +,- \rbrace} \mathcal{F}_{s,\scriptscriptstyle{\Lt},\epsilon}(\ell^2(\mathbbm{k})).
\end{equation}
According to \eqref{eq:h} and the mode decomposition for the Mukhanov-Sasaki and the tensor fields, it is convenient to introduce the mode-dependent frequencies, 
\begin{align} \label{eq:Def Frequencies}
\im{\omega}{\text{\Lms},\vec{k}}^{\scriptscriptstyle{(\Gv,\Gm,\KGv,\KGm)}} := \pp^2 \sqrt{\hbox{$\vec{k}$}^{\,2} + \mMS^2\,}, ~~~~ \im{\omega}{\text{\Lt},\vec{k}}^{\scriptscriptstyle{(\Gv,\Gm,\KGv)}} := \pp^2 \sqrt{18\,\hbox{$\vec{k}$}^{\,2} + 6\,(\pp\, \mT)^2\,}, 
\end{align}
Note that we set here and in the following, $\lambda\equiv1$, without loss of generality and in order to shorten the notation. We emphasize the parametric dependence on the homogeneous degrees of freedom expressed by the superscripts. Accordingly, the associated creation and annihilation operators for the modes encounter the same parameter dependence. We denote the pairs of annihilation and creation operators for the Mukhanov-Sasaki system for every mode $\vec{k} \in \mathbbm{k}$ as $(\aMS{\vec{k}}, \adMS{\vec{k}})$. Since the tensor perturbations decompose into the two polarizations, we introduce the short hand mode numbers $\vec{K} \in \mathbbm{K} := \lbrace \mathbbm{k},\epsilon \rbrace$ and write for the annihilation and creation operators $(\bT{\vec{K}},\bdT{\vec{K}})$. Both pairs of operators satisfy the standard commutation relations,
\begin{equation}
\im{[\,\aMS{\vec{k}}, \adMS{\vec{k}'}\,]}{\Lms} = \im{\delta}{\vec{k}, \vec{k}'}\,\Uf{\Fs{\Lms}}, ~~~ \im{[\,\bT{\vec{K}}, \bdT{\vec{K}'}\,]}{\Lt} =\im{\delta}{\vec{K}, \vec{K}'}\,\Uf{\Fs{\Lt,\epsilon}}.
\end{equation}
Here, the $\delta$ denotes the Kronecker delta. The suggested representation of the Hamilton constraint as a symbol function, i.e., as a function on the homogeneous phase space with values in the linear operators on $\Hf$, is given by,
\begin{align} 
\qf{h} &= L^3 \left(- \frac{1}{12} \frac{\Gm^2}{\Gv} + \frac{\KGm^2}{2\,\Gv^3} + \frac{1}{2}\,\varepsilon^2\,\mPhi^2 \Gv^3\,\KGv^2 + \Lambda\,\Gv^3 \!\right)\Uf{\Hf} + \frac{1}{\Gv} \sum_{\vec{k} \in \mathbbm{k}} \im{\omega}{\text{MS},\vec{k}}\; \adMS{\vec{k}}\,\aMS{\vec{k}} \otimes \Uf{\Fs{\Lt}} \nonumber \\
&~~~~+ \Uf{\Fs{\Lms}} \otimes \frac{1}{6\,\Gv} \sum_{\vec{K} \in \mathbbm{K}} \im{\omega}{\text{T},\vec{k}} \;\bdT{\vec{K}}{}\,\bT{\vec{K}}{} \nonumber \\
&=: \irm{E}{hom}^{\scriptscriptstyle{(\Gv,\Gm,\KGv,\KGm)}} \Uf{\Hf} + \frac{1}{\Gv} \sum_{\vec{k} \in \mathbbm{k}} \im{\omega}{\text{MS},\vec{k}}\; \adMS{\vec{k}}\,\aMS{\vec{k}} + \frac{1}{6\,\Gv} \sum_{\vec{K} \in \mathbbm{K}} \im{\omega}{\text{T},\vec{k}} \;\bdT{\vec{K}}\,\bT{\vec{K}} \label{eq:Hamilton Symbol}
\end{align}
where the standard normal ordering of operators has been induced. Furthermore, we omit the second respective first tensor factor if it is just a unity operator. \\

We develop the space adiabatic perturbation scheme in close analogy to the cosmological Klein-Gordon system in \cite{Schander Thiemann III} and we refer the reader to the latter work and the companion papers \cite{Schander Thiemann I,Neuser Schander Thiemann II} for a more detailed presentation of the scheme. Here, we only give the basic ideas in order to understand the given results of the formalism.

\section{Space Adiabatic Perturbation Scheme}

\subsection{The Parameter-Dependent Harmonic Oscillator}
The first step towards a rigorous application of space adiabatic perturbation theory is the solution of the parameter-dependent eigenvalue problem corresponding to \eqref{eq:Hamilton Symbol}. We define as a shorthand notation the set of homogeneous variables as, 
\begin{equation}
\qpb := (\Gv,\Gm,\KGv,\KGm).
\end{equation}
The eigenvalue problem in $\Hf$ is then formally given by,
\begin{equation} \label{eq:EVP}
\h{}{\qpb} \ef{\nD} = \Epd{n} \ef{\nD},
\end{equation}
where $\nD$ is a short form for the number of excitations $(\nMS{k},\nT{K})$ for every wave number $\vec{k}$, polarization $\epsilon$ and the degeneracy label of the Mukhanov-Sasaki system, $d \in \lbrace 1,...,D \rbrace$ and of the tensor system,  $d' \in \lbrace 1,..., D' \rbrace$. In particular, $\ef{\nD}$ is the eigensolution of the parameter-dependent Hamilton symbol $\h{}{\qpb}$ with the set of excitation numbers $(\nMS{k},\nT{K})$ and with energy,
\begin{equation} \label{eq:Epdn}
\Epd{n} = \irm{E}{hom}^{\qpb} + \frac{1}{\Gv} \sum_{\vec{k} \in \mathbbm{k}} \im{\omega}{\text{MS},\vec{k}}\; \nMS{k} + \frac{1}{6 \Gv} \sum_{\vec{K} \in \mathbbm{K}} \im{\omega}{\text{T},\vec{k}} \;\nT{K}.
\end{equation}
More precisely, every state in the Hilbert space $\Hf$  derives from the vacuum state $\Omega^{\qpb}$ by applying the desired number of creation operators $(\nMS{k},\;\nT{k}{+},\;\nT{k}{-})$ for every wave number $\vec{k}$. Space adiabatic perturbation theory chooses formally one such eigenstate which will be denoted by $\nD$. This choice is, however, arbitrary. In particular, at the end of the procedure, \emph{all} possible eigensolutions will be taken into account. The state as an excitation of the ground state is given by,
\begin{equation}
\ef{\nD} = \prod_{\vec{K} \in \mathbbm{K}} \frac{(\adMS{\vec{k}})^{\nMS{k}}}{\sqrt{\nMS{k}!}} \frac{(\bdT{\vec{K}})^{\nT{K}}}{\sqrt{\nT{K}!}}\,\Omega^{\qpb}.
\end{equation}
Even if we do not label the creation and annihilation operators explicitely by the homogeneous phase space variables, their dependence on them is crucial for the formalism. In particular, the formalism needs the derivatives of the eigenstates with respect to the homogeneous parameters which are to be understood as connections within the perturbative Hilbert space $\Hf$. Let therefore $x$ be one of the phase space variables $(\Gv,\Gm,\KGv,\KGm)$. We introduce the explicit representation of the Mukhanov-Sasaki wave function and the tensor wave functions as a product by,
\begin{equation}
\ef{\nD} =: \im{e}{\nd}^{\scriptscriptstyle{\Lms}} \cdot \prod_{\epsilon}  \im{e}{\ndp}^{\scriptscriptstyle{\Lt,\epsilon}}.
\end{equation} 
The derivative of $\ef{\nD}$ with respect to one of the homogeneous phase space variables $x$ is given by means of the connection symbol $\qf{\mathcal{A}}^{\scriptscriptstyle{x}} \in S(\Ps{\!\te{hom}},\mathcal{L}(\Hs{\te{p}}))$,
\begin{equation} \label{eq:Def Connection}
\frac{\partial \, \ef{\nD}}{\partial x} =: (\qf{\mathcal{A}}^{\scriptscriptstyle{x}})\,\ef{\nD} =: \sum_k (\mathcal{A}^{\scriptscriptstyle{x}})_{\nD}^{~\,k}\, \ef{k},~~~~~   (\mathcal{A}^{\scriptscriptstyle{x}})_{\nD}^{~\,k} \in \mathbb{R},~\forall \lbrace \nD, k \rbrace,
\end{equation}
where the summation over $k$ includes essentially \emph{all} possible excitation numbers within the Fock space $\Hs{\te{p}}$. However, there is only a countable number of $k$'s for which $(\mathcal{A}^{\scriptscriptstyle{x}})_{\nD}^{~\,k}$ is non-vanishing. In order to illustrate this, we introduce the phase space functions,
\begin{align}
\aone{\nMS{m}}{\Lms,x} &:= - f_{\Lms,\vec{m}}^{\,x} \frac{\sqrt{(\nMS{m}- 1)\cdot \nMS{m}}}{2}, \\
\atwo{\nMS{m}}{\Lms,x} &:= f_{\Lms,\vec{m}}^{\,x} \frac{\sqrt{(\nMS{m}+ 1)\cdot (\nMS{m} +2)}}{2}, \\
\aone{\nT{M}}{\Lt,x} &:= - f_{\Lt,\vec{m}}^{\,x} \frac{\sqrt{(\nT{M}- 1)\cdot \nT{M}}}{2}, \\
\atwo{\nT{M}}{\Lt,x} &:= f_{\Lt,\vec{m}}^{\,x} \frac{\sqrt{(\nT{M}+ 1)\cdot (\nT{M} +2)}}{2},
\end{align}
where the functions $f$ derive from the frequencies, \eqref{eq:Def Frequencies},
\begin{equation}
f_{\Lms,\vec{m}}^{\,x} = - \frac{\pp^4\,\partial_x \mMS^2}{4\, \im{\omega}{\text{MS},\vec{m}}^2}, ~~~~ f_{\Lt,\vec{M}}^{\,x} = - \frac{3\,\pp^6\,\partial_x \mT^2}{2\, \im{\omega}{\text{T},\vec{M}}^2}
\end{equation}
The derivative of $\ef{\nD}$ with respect to $x$ is then given by,
\begin{align} \label{eq:Derivative States}
\frac{\partial \, \ef{\nD}}{\partial x} =& \sum_{\vec{m} \in \mathbbm{k}} \left( \aone{\nMS{m}}{\Lms,x} \eMSs{-2} + \atwo{\nMS{m}}{\Lms,x} \eMSs{+2} \right) \prod_{\epsilon}  \im{e}{\ndp}^{\scriptscriptstyle{\Lt,\epsilon}} \\
& +  e_{\nd}^{\scriptscriptstyle{\Lms}} \prod_{\epsilon} \sum_{\vec{m} \in \mathbbm{k}} \left( \aone{\nT{M}}{\Lt,x} \eTs{-2} + \atwo{\nT{M}}{\Lms,x} \eTs{+2} \right) \im{e}{\ndp}^{\scriptscriptstyle{\Lt,\epsilon' \neq \epsilon}}. \nonumber
\end{align}
This simply corresponds to a mode shifting by plus or minus two, seperately for every single mode of the given eigenstate, multiplying the shifted state by a frequency-dependent and mode number- dependent function $\alpha$ and eventually summing up all shifted states.  Note also that the connections here are real-valued since the perturbation wave functions are real-valued, too. \\

Equipped with these wave function in the fast Hilbert space $\Hf$, it is now straightforward to define a projection operator associated to the chosen quantum state with quantum number(s) $\nD$. Using the inner product, $\im{\langle \cdot, \cdot \rangle}{\Hf}: \Hf \times \Hf \rightarrow \mathbb{C}$ of the Fock spaces, this projection operator is defined as,
\begin{equation} \label{eq:Moyal Projector 0}
\pipd{0} = \sum_{\delta} \ef{\nD} \,\langle \ef{\nD} , \cdot \rangle.
\end{equation}
The index `$0$' indicates that this orthogonal projection operator represents the zeroth order iteration step of our constructive space adiabatic perturbation scheme.

\subsection{Structural Ingredients}
Space adiabatic perturbation theory requires three structural ingredients in order to be applicable. We refer to the companion papers, \cite{Schander Thiemann I,Neuser Schander Thiemann II,Schander Thiemann III}, in which we discuss these requirements for similar models in detail. In brief, these requirements are that
\begin{enumerate}
\item first, the quantum Hilbert space of the system decomposes as a tensor product of the form, $\Hi = L^2(\mathbb{R}^h) \otimes \Hs{\te{p}}$, where $L^2(\mathbb{R}^h)=\Hs{\te{hom}}$ is the Hilbert space associated with the $h$ homogeneous degrees of freedom in our context, 
\item second, deformation quantization with the Weyl ordering applies to the homogeneous part of the system, and in particular, the quantum Hamilton constraint of the model is the Weyl quantization of a semiclassical symbol $\qf{h} \in S(\pp;\Ps{\!\te{hom}}, \mathcal{L}(\Hs{\te{p}})) $, 
\item third, the Hamilton symbol $\qf{h}^{(q,p)}$ has a pointwise isolated part within spectrum $\sigma^{(q,p)}$. ``Pointwise'' means here that for every point $(q,p) \in \Ps{\!\te{hom}}$, the part of the spectrum  $\im{\sigma}{\nD}^{(q,p)}$, associated with some quantum number $\nD$ of the perturbation system is isolated from the remainder of the spectrum $\sigma^{(q,p)} \setminus \im{\sigma}{\nD}^{(q,p)}$.
\end{enumerate} 
Given these requirements, the scheme decomposes into three steps,
\begin{enumerate}
\item the construction of the Moyal projector $\qf{\pi} \in S(\pp;\Ps{\!\te{hom}}, \mathcal{L}(\Hs{\te{p}}))$,
\item the construction of the Moyal unitary $\qf{u} \in S(\pp;\Ps{\!\te{hom}}, \mathcal{L}(\Hs{\te{p}}))$,
\item the construction of the effective Hamiltonian, $\heff{}{} \in S(\pp;\Ps{\!\te{hom}}, \mathcal{L}(\Hs{\te{p}}))$.
\end{enumerate}
The symbol classes $S(\pp;\Ps{\!\te{hom}}, \mathcal{L}(\Hs{\te{p}}))$ comprise the operator-valued symbols which have the form of a formal power series in the parameter $\pp$ and they are denoted as ``semiclassical symbols''. For more details about the symbol classes, see \cite{Teufel} and references therein.

\subsection{The Moyal Projector} \label{sec:Moyal Proj}

The space adiabatic perturbation scheme starts with the construction of a semiclassical symbol, $\pipd{}:\Ps{\!\te{hom}} \rightarrow S(\pp;\Ps{\!\te{hom}}, \mathcal{L}(\Hs{\te{p}}))$ whose Weyl quantization is a spectral projector on the full Hilbert space $\mathcal{H}$. As indicated before, the construction proceeds iteratively with respect to the space adiabatic perturbation parameter $\pp$ such that it produces a set of symbols, $\lbrace \pipd{\nD,N} \rbrace_{N \in \mathbb{N}}$. The full Moyal projector symbol has the form of a formal power series in $\pp$,
\begin{equation}
\pipd{n} = \sum_{\delta} \sum_{N=0}^{\infty} \pp^N \pipd{\nD,N}, ~~~ \pipd{\nD,N} \in \mathcal{S}(\Ps{\!\te{hom}},\mathcal{L}(\Hs{\te{p}})).
\end{equation}
The sum over $\delta$ indicates the degeneracy of the chosen subspace with formal quantum number $n$, which is of course the set of excitation numbers with respect to the Mukhanov-Sasaki and the tensor system. We omit the index of the chosen subspace in the following in order to simplify the notation. \\
We introduce the Moyal product ``\,$\im{\star}{\pp}$\!'' which is the star product for the Weyl ordering i.e., the pull back of the operator Weyl ordered multiplication on the space of semiclassical symbols. Thereby, the iterative construction of the Moyal projector symbol of order $N$ rests on the requirements, that
\begin{enumerate}
\item[1)] $\im{\qf{\pi}}{(N)}\, \im{\star}{\pp}\, \im{\qf{\pi}}{(N)} - \im{\qf{\pi}}{(N)} = \mathcal{O}(\pp^{N+1})$,
\item[2)] $\im{\qf{\pi}}{(N)}^{\ast}- \im{\qf{\pi}}{(N)} = \mathcal{O}(\pp^{N+1})$,
\item[3)] $\im{\qf{\pi}}{(N)} \, \im{\star}{\pp}\, \qf{h} - \qf{h} \, \im{\star}{\pp}\, \im{\qf{\pi}}{(N)} = \mathcal{O}(\pp^{N+1})$, 
\end{enumerate}
which ensures that the Moyal projector has indeed the properties of a projector up to the considered order in $\pp$, and that it commutes with the Hamiltonian symbol, also up to the considered order. Once more, we emphasize that the scheme is constructive i.e., the Moyal projector symbol at every desired order reduces in principle to the zeroth order Moyal projector symbol $\im{\qf{\pi}}{0}$ and derivatives thereof. The same holds true for the Moyal unitary which can be reduced to the zeroth order symbol $\us{0}{}$ and its respective derivatives. Most importantly, also the formula for the effective Hamilton symbol $\heff{}{}$ reduces to $\im{\qf{\pi}}{0}$ and $\us{0}{}$ and their derivatives. Hence, it suffices in principle to give the respective formula by means of the symbols $\im{\qf{\pi}}{0}$ and $\us{0}{}$. However, it is very instructive to present the explicit Moyal projector and the Moyal unitary at higher orders so as to provide physical insights into the space adiabatic scheme. \\

In this work, we restrict our interest to the construction of the effective Hamilton symbol up to second order in $\pp$. It can be shown that the construction scheme requires only the Moyal projector symbol up to \emph{first} order as well as the Moyal unitary symbol up to \emph{first} order. For our purposes, it will thus suffice to explicitely compute $\im{\qf{\pi}}{(1)}$ and $\us{(1)}{}$. \\  

The starting point for the construction of $\im{\qf{\pi}}{(1)}$ is the zeroth order Moyal projector symbol in \eqref{eq:Moyal Projector 0}. Then, condition $1)$ gives a result for the diagonal contributions of $\im{\qf{\pi}}{1}$, while condition $3)$ defines the off-diagonal terms of $\im{\qf{\pi}}{1}$. The only necessary terms of $\pli{1}$ for the computation of the effective Hamiltonian up to second order are the off-diagonal contributions, $\pli{1}^{\te{OD}}$. In the appendix, we show that $\pli{1}^{\te{OD}}$ involves the Poisson brackets with respect to the canonical pair $(\KGv, \KGm)$. Since these are the only Poisson brackets involved in our computations, we will simply denote them as $\lbrace \cdot, \cdot \rbrace$. For our model, the only remaining terms are then, 
\begin{align}
\im{\qf{\pi}}{1}^{\scriptscriptstyle{\te{OD}}} =& - \frac{i}{2} \left( \qf{h}- \!\im{E}{n} \right)^{-1} \!\left( \Uf{\Hf}\! \!- \im{\qf{\pi}}{0} \right) \!\lbrace \qf{h} + \im{E}{n}, \im{\qf{\pi}}{0} \rbrace \,\im{\qf{\pi}}{0} - \frac{i}{2} \, \im{\qf{\pi}}{0} \im{\lbrace \qf{\pi}}{0}, \qf{h} + \im{E}{n} \rbrace \left( \qf{h}-\! \im{E}{n} \right)^{-1}\! \left( \Uf{\Hf} \! \!- \im{\qf{\pi}}{0} \right). \label{eq:Pi1}
\end{align}
Here, $\qf{h}$ is the full Hamiltonian symbol from \eqref{eq:h}, $\im{E}{n}$ is the $(q,p)$-dependent energy of the relevant subspace with quantum number $n$, given in \eqref{eq:Epdn} The operator $\left( \qf{h}- \im{E}{n} \right)^{-1} \left( \Uf{\Hf}\! - \im{\qf{\pi}}{0} \right)$ is the zeroth order contribution of the local Moyal resolvent of $\qf{h}$ at $\im{E}{n}$. \\

In the appendix, we give the explicit expression for $\pli{1}^{\te{OD}}$ for our model in terms of the connections defined in \eqref{eq:Def Connection}.

\subsection{The Moyal Unitary} \label{sec:Moyal Unitary}
The Weyl quantization of the Moyal projector in the preceding section projects on a subspace of $\Hi$ which depends on $\pp$ in a non-trival way. The dynamics within this subspace are difficult to extract. It is thus reasonable to unitarily map the corresponding subspace to a simpler reference space $\mathcal{K}$. Here, the simplest and physically most convenient choice for a reference space is the perturbation subspace associated with one particular point in the homogeneous phase space, $(q,p) \equiv (\im{q}{0},\im{p}{0})$. We denote the reference projector associated with $\im{\qf{\pi}}{0}$ as,
\begin{equation}
\pli{\te{R}} := \sum_{\delta} \efzero{\nD}{} \,\langle \efzero{\nD}{} , \cdot \rangle.
\end{equation}
In order to connect the basis states associated with the eigenvalue problem in \eqref{eq:EVP} for arbitrary $(q,p)$ with the ones in $\mathcal{K}$, we define the ``Moyal unitary'' at zeroth order,
\begin{align} \label{eq:Moyal unitary 0}
\us{0}{} = \sum_{\im{m}{\rho}} \efzero{\im{m}{\rho}} \,\langle \ef{\im{m}{\rho}} , \cdot \rangle.
\end{align}
In fact, $\us{0}{}$ maps $\pli{0}$ to $\pli{\te{R}} $ in the sense that, $\pli{\te{R}}  = \us{0}{}\cdot\pli{0}\cdot\us{0}{\ast}$. Furthermore, it is unitary with respect to the operator product in $\Hs{\te{p}}$, in particular, $\us{0}{}\cdot\us{0}{\ast} = \Uf{\Hs{\te{p}}}$. Again, we aim at constructing iteratively the first contributions to a Moyal unitary which is a formal power series with respect to $\pp$,
\begin{equation}
\us{\nD}{(q,p)} = \sum_{N=0}^{\infty} \pp^N \us{\nD,N}{(q,p)}, ~~~ \us{\nD,N}{(q,p)} \in S(\Ps{\!\te{hom}},\mathcal{L}(\Hs{\te{p}})).
\end{equation} 
It is instructive to pronounce the following requirements for the construction of the $\us{\nD,N}{(q,p)}$'s,
\begin{enumerate}
\item[1)] $\us{(N)}{}\,\im{\star}{\pp}\,\qf{\pi}\,\im{\star}{\pp}\,\us{(N)}{\ast} - \irm{\qf{\pi}}{R} = \mathcal{O}(\pp^{N+1})$
\item[2)] $\us{(N)}{}\,\im{\star}{\pp}\, \us{(N)}{\ast} - \Uf{\Hs{\te{p}}} = \mathcal{O}(\pp^{N+1})$
\item[3)] $ \us{(N)}{\ast}\,\im{\star}{\pp}\, \us{(N)}{} - \Uf{\Hs{\te{p}}} = \mathcal{O}(\pp^{N+1})$
\end{enumerate}
These conditions give an expression for $\us{1}{}$ after inserting the zeroth order Moyal unitary from \eqref{eq:Moyal unitary 0}. Several contributions to $\us{1}{}$ vanish as it will be shown in the appendix, and the only remaining term is, 
\begin{align}
\im{\qf{u}}{1} = \left[\irm{\qf{\pi}}{R}, \im{\qf{u}}{0}\,\im{\qf{\pi}}{1}^{\te{OD}} \im{\qf{u}}{0}^{\ast} \right]\,\im{\qf{u}}{0}. 
\end{align}
The brackets, $\left[\cdot,\cdot \right]$\,, denote the commutator brackets with respect to the perturbation operators. Again, the explicit expression of $\us{1}{}$ for our model will be given in the appendix.  

\subsection{The Effective Hamiltonian} \label{sec:Eff Ham}
The intent of space adiabatic perturbation theory is to construct an effective Hamiltonian symbol which encodes the back reactions from the perturbation fields onto the  homogeneous background degrees of freedom. Thus, the las step consists in pulling the dynamics of the chosen subspace to the $\pp$-independent subspace $\mathcal{K}$. In this regard, the effective Hamiltonian is defined as,
\begin{equation} \label{eq:heff def}
\heff{}{} := \us{}{}\,\im{\star}{\pp}\,\qf{h}\,\im{\star}{\pp}\,\us{}{\ast}.
\end{equation}
Its Weyl quantization, $\irm{\qs{\qf{h}}}{eff}$, has the properties,
\begin{align}
\left[ \irm{\qs{\qf{h}}}{eff}, \im{\qs{\qf{\pi}}}{\te{R}} \right] &= 0, \\
e^{-i \qs{\qf{h}} s} - \qs{\us{}{\ast}} e^{-i \irm{\qs{\qf{h}}}{eff} s} \qs{\us{}{}} &= \mathcal{O}(\pp^{\infty} |s|).
\end{align}
where $s\in \mathbb{R}$. \\

We construct $\heff{}{}$ perturbatively by means of \eqref{eq:heff def} up to second order. As before, we assume a formal power series in $\pp$ for the generic form of the semiclassical symbol $\heff{}{}$,
\begin{equation}
\heff{}{(q,p)} = \sum_{N=0}^{\infty} \pp^N \heff{,N}{(q,p)},~~~  \heff{,N}{(q,p)} \in S(\Ps{\!\te{hom}},\mathcal{L}(\Hs{\te{p}}))
\end{equation}
The effective Hamiltonian at zeroth order accounts for the diagonal energy contributions within every possible subspace of the full Hilbert space associated with $\im{m}{\rho}$ which becomes rotated to the reference subspace $\mathcal{K}$, 
\begin{align}
\heffscp{,0} = \sum_{\im{m}{\rho}} \Epd{m} \efzero{\im{m}{\rho}} \,\langle \efzero{\im{m}{\rho}} , \cdot \rangle.
\end{align}
We restrict our interest to the relevant subspace, which yields,
\begin{align} \label{eq:heff 0}
\heffscp{,0,\text{r}} = \irm{\qf{\pi}}{R}\cdot\heffscp{,0}\cdot \irm{\qf{\pi}}{R} = \Epd{n} \,\pli{\te{R}},
\end{align}
and we recall that $\nD$ is one specific, but arbitrary, set of quantum numbers for the perturbation subsystem.
The result corresponds to the Born-Oppenheimer limit of quasi-adiabatic systems. Note that the homogeneous degrees of freedom already encounter backreactions from the perturbations by means of the ``bare'' energy from the relevant subspace with quantum number $\nD$.\\

The first order effective Hamiltonian symbol, $\heff{,1}{}$ derives from \eqref{eq:heff def} with our result of the zeroth order contribution, \eqref{eq:heff 0}. Symbolically, it is given by,
\begin{equation} \label{eq:heff1}
\heffscp{,1} = \left( -\, \heff{,0}{}\,\im{\qf{u}}{1}  + \im{\qf{u}}{1}\, \qf{h} + \frac{i}{2} \im{\lbrace \im{\qf{u}}{0}, \qf{h} \rbrace}{\phi} - \frac{i}{2}  \im{\lbrace \heff{,0}{} , \im{\qf{u}}{0} \rbrace}{\phi}\right) \im{\qf{u}}{0}^{\ast}.
\end{equation}
Again, our focus is on the effective Hamiltonian symbol for the relevant subspace. We obtain that this symbol vanishes,
\begin{align}
\heffscp{,1,\text{r}} = \irm{\qf{\pi}}{R}\cdot\heffscp{,1}\cdot \irm{\qf{\pi}}{R} = 0,
\end{align}
as we will show in the appendix. \\

Finally, we compute the second order effective Hamilton symbol, $\heffscp{,2,r}$ for our model. The only non-vanishing contributions are given in terms of the preceding symbols by, 
\begin{align} \label{eq:heff 2}
\heffscp{,2,\text{r}} =&\, \frac{1}{8}\, \pli{\te{R}} \left(- \frac{\partial^2 \im{\qf{u}}{0}}{\partial \KGv^2}\, \frac{\partial^2(\qf{h}-\!\im{E}{n})}{\partial \KGm^2} + 2\,\frac{\partial^2 \us{0}{}}{\partial \KGv \, \partial \KGm} \frac{\partial^2 (\qf{h}-\! \im{E}{n})}{\partial \KGm\,\partial \KGv} - \frac{\partial^2 \us{0}{}}{\partial \KGm^2} \, \frac{\partial^2 (\qf{h}-\!\im{E}{n})}{\partial \KGv^2} \right) \im{\qf{u}}{0}^{\ast}\,\irm{\qf{\pi}}{R}. \nonumber \\
 & +\frac{i}{2}\, \pli{\te{R}}\, \lbrace \us{1}{}, \qf{h} + \im{E}{n} \rbrace \, \us{0}{\ast}\,\pli{\te{R}}.
\end{align}
We provide the explicit expression of $\heffscp{,2,\text{r}}$ in terms of the connections in the appendix. As it turns out, there is only one remaining contribution for the backreactions, and the complete effective Hamiltonian symbol up to second order in $\pp$ is given by, 
\begin{align} \label{eq:heff 2}
\heffscp{,(2),\text{r}} = \Epd{n} \,\pli{\te{R}} +  \pp^2\,\sum_{\delta} \efzero{\nD} \,\langle \efzero{\nD} , \cdot \rangle \sum_{\vec{m}} \frac{1}{\left( \vec{m}^2 + \mMS^2 \right)^{5/2}} \left(2\,\nMS{m} +1\right)\, \mathit{h}^{\scriptscriptstyle{(q,p)}},
\end{align}
where we defined the phase space function,
\begin{align}   \label{eq:heff 2 function}
\mathit{h}^{(q,p)}:= - \frac{9}{2}\,\mPhi^4 \frac{\KGm^4}{\Gv^3 \Gm^2} 
\end{align}
Due to the high inverse power of the frequency-factor in $\heffscp{,2,\text{r}}$, the second order effective Hamiltonian contributions converge, even if there are terms which contribute for every possible wave number and not only for the finite number of excited states. The function $\mathit{h}$ gives the explicit form of the effective Hamiltonian at second order. \\

Note that the effective Hamiltonian includes the rescaled variables $\Gm$ and $\KGm$ in \label{eq:Scaling hom} and has furthermore been rescaled by a factor $\pp^2$. We express our result in the standard cosmological variables, multiply the constraint by $\pp^{-2}$, and we obtain,
\begin{align}
\heffscp{,(2),\text{r}} =& \sum_{\delta} \efzero{\nD} \,\langle \efzero{\nD} , \cdot \rangle \nonumber \\
& \left\lbrace L^3 \left(- \frac{1}{12} \frac{\pp^2 \Gm^2}{a} + \frac{\KGm^2}{2 \,\Gv^3} + \frac{1}{2} \mPhi^2 \Gv^3 \KGv^2 + \frac{\Lambda}{\pp^2} \Gv^3 \right) \right. \label{eq:result1} \\
&~~ + \frac{1}{a} \sum_{\vec{m} \in \mathbbm{k}} \sqrt{\hbox{$\vec{m}$}^{\,2} + \mMS^2\,}\,\nMS{m}  + \frac{1}{6\,a} \sum_{\vec{M} \in \mathbbm{K}} \sqrt{18\,\hbox{$\vec{M}$}^{\,2} + 6\,(\pp\, \mT)^2\,}\,\nT{M} \label{eq:result2}\\
&~~ -  \left. \sum_{\vec{m} \in\mathbbm{k}} \frac{1}{\left( \vec{m}^2 + \mMS^2 \right)^{5/2}} \left(2\,\nMS{m} +1\right)\,\frac{9}{2}\,\mPhi^4\,\frac{\KGm^4}{\Gv^3 \Gm^2} \right\rbrace, \label{eq:result3}
\end{align}
where also the tensor and Mukhanov-Sasaki masses need to be expressed in the original variables. 
We emphazise that the contributions \eqref{eq:result1}, \eqref{eq:result2} represent 
the standard Born-Oppenheimer effective Hamiltonian, while \eqref{eq:result3} results 
from the improved analysis with the space adiabatic scheme. Note also that for the truncation 
at second order, no tensor backreactions, except for the trivial Born-Oppenheimer contribution, occur.

Note that even if the vacuum energy band is considered (i.e. the excitation numbers
$n_\ast$ vanish) there is a non-vanishing contribution of the {\it Casimir type}.

\section{Positive Mass Restricted Model}

\subsection{Positive Mass Squared Issue}

The Mukhanov-Sasaki and graviton mass squared terms are not manifestly positive in all regions of the phase space of the homogeneous degrees of freedom. In order to avoid the instabilities accompanied by tachyon fields, one has essentially two options, as stated 
in the companion papers. Either one reduces the Fock space by hand to those modes for which the frequency squared function is not negative or one restricts the phase space by hand to its positive mass squared region which arises as the image under an embedding of an unrestricted phase space. The first option has the 
disadvantage that the framework of space adiabatic perturbation theory has to be modified in the sense that for 
each point $(q,p)$ in the homogeneous phase space, we must drop those modes $\vec{k}$ from the construction of the Moyal projections and unitarities respectively for which the frequency squared function is negative which means that the unitarity turns into a partial isometry and leads to major modifications of the framework. This means that we declare the energy band eigenvalues to vanish for different
modes thus violating the gap condition. The second option avoids these complications, but the restriction of the phase space is also a priori unjustified and needs to be revisited. To partly justify it, note that at least the positive mass region contains the kernel of the homogeneous part of the 
classical constraint so that one restricts the phase space to an open neighbourhood of the purely homogeneous constraint surface. The logically possible third option, namely to allow the negative mass region requires to depart from Fock representations there which would result in even more drastic modifications of the entire framework.\\ 

In what follows we consider the second option and restrict for simplicity to the case of vanishing inflaton potential and zero cosmological constant $\Lambda=V=0$, a case often considered in Loop Quantum Cosmology. \\
Therefore, consider the scaled homogeneous variables $(\Gv, \Gm\,\KGv, \KGm)$ with Poisson brackets, $\{\Gv,\Gm\}=\pp^2/L^3,$ and $\{\KGv,\KGm\}=\pp/L^3$. In that case, recalling the adiabatic parameter $\pp^2=\frac{\kappa}{\lambda}$, the Mukhanov-Sasaki and graviton mass squared functions read respectively,
\begin{align}
\mMS^2 &= - \frac{1}{18} \frac{\Gm^2}{\Gv^2} + \frac{7\,\KGm^2}{2\,\Gv^4} - 18 \frac{\KGm^4}{\Gv^2\,\Gm^2}, ~~~ (\pp\,\mT)^2 =\frac{\Gm^2}{6\,\Gv^2}
\end{align}
Note that because of $V=0$, the variable $\KGv$ is cyclic. Evidently, $(\pp\,\mT)^2\ge 0$ manifestly, however, not $\mMS^2$. After some algebraic manipulations we can write $\mMS^2$ as a manifestly positive quantity. Therefore, we define, $y:= \Gv\,\Gm$, and we write,
\begin{align}
\mMS^2=\frac{18}{\Gv^4 y^2}\;[\,c_+ y^2-\KGm^2]\;[\,\KGm^2-c_- y^2],
\end{align}
where,
\begin{align}
c_\pm^2=\frac{1}{72}\left(7\pm \sqrt{33} \right)
\end{align}
Note that $c_+ >1>c_- >0$. If $\mMS^2>0$ we must constrain $\KGm^2$ by, 
\begin{align}
c_+^2 y^2 > \KGm^2 > c_-^2 y^2
\end{align}
which is solved by the explicit parametrisation
\begin{align}
\KGm=y\; \KGnv
\end{align}
with $\KGnv\in [-c_+,-c_-]\cup [c_-,c_+]$. We note that $\Gv\,\Gm$ is conjugate to $\alpha:= \ln a$. Furthermore, we introduce the variables,  
\begin{align}
\KGnv := \,\frac{\KGm}{y},~~~ \KGnm := -\,y\,\KGv.
\end{align}
By this parametrisation, the symplectic structure can be pulled back. Dropping total differentials, we obtain,
\begin{align}
\frac{\pp^2}{L^3}\,\theta = - \left(\Gv\,\mathrm{d}\Gm + \pp\,\KGv\,\mathrm{d}\KGm \right) = - \left( (\alpha + \pp\,\KGv\,\KGnv)\,\mathrm{d}y - \pp\,\KGnm\,\mathrm{d}\KGnv \right). 
\end{align}
It is manifest to identify $\KGnm$ as a new momentum variable and $\KGnv$ as its conjugate variable. Similarly, $y$ can serve as a new momentum variable with conjugate variable $(\alpha + \pp\,\KGv\,\KGnv)$. \\
In a final step, it is useful to introduce another canonical transformation. Therefore, we define as a canonical variable,
\begin{align}
\Gnv := \exp (\alpha + \pp\,\KGv\,\KGnv) = \Gv \cdot \exp \left( \pp\, \frac{\KGv\,\KGm}{\Gv\,\Gm} \right). 
\end{align}
The variable $\Gnm$, with $y =: \Gnv\,\Gnm$ serves as a conjugate momentum for $\Gnv$, and in terms of the initial variables, it reads, 
\begin{align}
\Gnm &= \Gm \cdot \exp \left(-\,\pp\, \frac{\KGv\,\KGm}{\Gv\,\Gm} \right).
\end{align}
Note also that the following identity holds, 
\begin{align}
\pp\,\frac{\KGv\, \KGm}{\Gv\,\Gm} = - \pp\, \frac{\KGnv \,\KGnm}{\Gnv\,\Gnm}.
\end{align}
Finally, the transformation $T: \mathbb{R}^4 \supset U \ni (\tilde{q},\tilde{p}) \rightarrow \qpb \in W \subset \mathbb{R}^4$, with $U,W$ of $\mathbb{R}^4$, which maps the new variables canonically on the initial ones, is given by,
\begin{align}
\Gv &= \Gnv\cdot \exp \left( \pp\, \frac{\KGnv \,\KGnm}{\Gnv\,\Gnm} \right), \label{eq:T1}\\
\Gm &= \Gnm \cdot \exp \left(-\,\pp\, \frac{\KGnv\,\KGnm}{\Gnv\,\Gnm} \right), \\
\KGv &= - \frac{\KGnm}{\Gnv\,\Gnm}\\
\KGm &= \Gnv\,\Gnm\,\KGnv. \label{eq:T2}
\end{align}
We emphasize that in the new space adiabatic perturbation scheme, we can treat $\Gnm$ like $\Gm$ with rescaling $\pp^2$ and $\KGnm$ as $\KGm$ with rescaling $\pp$. In the new variables, the homogeneous part of the Hamilton constraint and the masses, $\mMS^2$, $(\pp\,\mT)^2$ are given by,
\begin{align}
\irm{\tilde{E}}{hom} &= \frac{L^3}{2} \frac{(\Gnm)^2}{b}  \exp \left(- \pp\, \frac{\KGnv\,\KGnm}{\Gnv\,\Gnm} \right)\left( \KGnv^2 - \frac{1}{6} \right),\\
\mMSt^2 &= 18\,\frac{(\Gnm)^2}{\Gnv^2}\,\exp \left(-4\,\pp\, \frac{\KGnv\,\KGnm}{\Gnv\,\Gnm} \right) \left(c_+ - \KGnv^2 \right)\left(\KGnv^2-c_- \right), \label{eq:mMS}\\
(\pp\, \mTt)^2 &= \frac{1}{6}\,\frac{(\Gnm)^2}{\Gnv^2}\,\exp \left(-4\,\pp\, \frac{\KGnv\,\KGnm}{\Gnv\,\Gnm} \right). \label{eq:mT}
\end{align}
Now all mass terms are manifestly positive and it is consistent to quantise the phase space $T^\ast I$ where $I$ is the union of the two intervals defined by $c_\pm$.\\
\\
The space adiabatic perturbation scheme is now to be carried out in terms of the slow rescaled momenta $\KGnm$ and $\Gnm$. In the previous section we carried it out in terms of $\KGm$ and $\Gm$. The calculations are given in the next section. Fortunately, the dominant contribution to the Moyal product at the second adiabatic order of interest here, comes from the first term which is nothing but a Poisson bracket when acting on products of functions and since the transition from $(\Gv,\Gm),(\KGv,\KGm)$ to $(\Gnv,\Gnm),(\KGnv, \KGnm)$ preserves those, the computations carry over in a direct way.\\

To quantise $\Gv$ in the resulting effective Hamiltonian which depends non-polynomially on all four degrees of freedom $(\Gnv,\Gnm,\KGnv,\KGnm)$ in the combination $\Gnv\,\exp(\pp\,\KGnv\,\KGnm/(\Gnv \Gnm))$, according to the space adiabatic scheme, we expand the exponential to the desired order in $\pp$ which yields an expression which depends only on positive powers of $\KGnv$ and $\KGnm$ and also on negative powers of finite order of $\Gnv$ and $\Gnm$. The resulting expression must then be Weyl ordered. However, we first have to bring the phase space into the form of (copies of) a cotangent bundle over the real line in order that the Fourier transforms in terms of which the Weyl quantisation is performed can be applied (alternatively one can develop Weyl quantisation
on $T^\ast I$ independently).

Concerning the quantisation of $T^\ast I$ where $I=I_1\cup..\cup I_n$ is a union of 
disjoint intervals $I_k$ we note that the Hilbert space $L_2(I,\mathrm{d}\KGnv)$ of square integrable
functions $\psi$ over $I$ is specified uniquely by the restrictions $\psi_k=\psi_{|I_k}$
which shows that $L_2(I,\mathrm{d}\KGnv)=\oplus_k L_2(I_k,\mathrm{d}\KGnv)$. Now each $I_k$ is of the form  
$[a,b]$ and using suitable maps, e.g.  $\KGnv=f(x)=a+(1+2{\rm arctan}(x)/\pi)(b-a)/2$ 
with $df/dx(x)>0$ and 
and the associated conjugate momenta $y:=u\; df/dx(x)$ i.e. $\KGnm=y/(df/dx)(x)$ we 
may think of $T^\ast I_k$ as $T^\ast \mathbb{R}$. We pick the Hilbert space 
$L_2([a,b],\mathrm{d} \KGnv)$ on which $\KGnv$ acts by multiplication and $u$ as $i d/dw$ (subject
to boundary conditions to make it self-adjoint). We consider
then the symmetric operators $X:=f^{-1}(\KGnv),\; Y=\sqrt{f'(x)}_{x=f^{-1}(w)}
u \sqrt{f'(x)}_{x=f^{-1}(w)}$ which satisfy the canonical commutation relations 
$[Y,X]=i$ and the unitary map 
$$
U: L_2([a,b],dw)\to L_2(\mathbb{R},dx);\;
(U\psi)(x):=\sqrt{f'(x)} \psi(f(x))=\hat{\psi}(x)
$$
with inverse $(U^{-1} \hat{\psi})(\KGnv)=(\hat{\psi}/\sqrt{f'})(f^{-1}(\KGnv)$ then one may 
check that $\hat{X}=U X U^{-1}$ acts by multiplication by $x$ and 
$\hat{Y}=U Y U^{-1}$ by $i d/dx$. Now in every symbol we express $\KGnm$ and $\KGnv$ in terms of 
$x,y$ and use their Weyl quantisation for $\hat{X}, \hat{Y}$ on $L_2(\mathbb{R},dx)$. 
Using 
above formulae, the result may then be translated back in terms of $\KGnm,\KGnv, L_2([a,b],\mathrm{d}\KGnv)$.
For instance the Weyl quantisation of $\KGnm \KGnv=f(x)/f'(x) y$ on $L_2(\mathbb{R},dx)$ 
yields
\begin{eqnarray}*
&&\frac{1}{2}[f(\hat{X})/f'(\hat{X}) \hat{Y}+\hat{Y}f(\hat{X})/f'(\hat{X})]
\nonumber\\
&=& \frac{1}{2} U[f(X)/f'(X) Y+Y f(X)/f'(X)]U^{-1}
\nonumber\\
&=& \frac{1}{2}
U[w/\sqrt{f'(f^{-1}(\KGnv))} \KGnm \sqrt{f'(f^{-1}(\KGnv))}+\sqrt{f'(f^{-1}(\KGnv))} \KGnm/\sqrt{f'(f^{-1}(\KGnv))} \KGnv]U^{-1}
\nonumber\\
&=& \frac{1}{2}(\KGnv \KGnm+\KGnm \KGnv+ \KGnv/\sqrt{f'(f^{-1}(w))}[\KGnm,\sqrt{f'(f^{-1}(\KGnv))}]-[\KGnv,\sqrt{f'(f^{-1}(w))}/\sqrt{f'(f^{-1}(\KGnv))} \KGnv] U^{-1}
\nonumber\\
&=& U\frac{1}{2}(\KGnm \KGnv+ \KGnv \KGnm) U^{-1}
\end{eqnarray}   
which shows that Weyl quantisation of $uw$ on $L_2([a,b],\mathrm{d}\KGnv)$ yields the expected symmetric
result.

Concluding, we simply have to rewrite the formulae of the previous section given there in terms of $(\Gv,\Gm,\KGv,\KGm)$ now in terms of $(\Gnv,\Gnm,\KGnv,\KGnm)$ and symmetrically order the outcome of the space adiabatic perturbation analysis 
as if $(\KGnv,\KGnm)$ would take values in $\mathbb{R}$. The quantisation of $(\KGnv,\KGnm)$ takes place
on $L_2([-c_+,-c_-],\mathrm{d}\KGnv)\oplus L_2([c_-, c_+])$.

\subsection{Results for Manifestly Positive Mass Squared
Variables}
We apply the space adiabatic perturbation scheme to the model with canonically transformed variables $(\Gnv, \Gnm,\KGnv, \KGnm)$. The scheme proceeds in the very same manner. However, it must be taken into account that the Moyal product transforms according to the canonical transformation in the preceding section, \cite{Blaszak Domanski}. We note, that the canonical transformation preserves the Moyal product for the first non-trival product contributions, i.e., for the Moyal product with respect to $(\KGv,\KGm)$ at first order in $\pp$ and for the Moyal product with respect to $(\Gv,\Gm)$ at second order in $\pp$. This is because the first non-trivial contributions are simply the Poisson brackets with respect to the canonical variables, which are preserved under canonical transformations. However, for the second order Moyal product, which is (in principle) of interest for our perturbation scheme, the Moyal product generates additional terms. Let therefore $T$ again be the canonical transformation from \eqref{eq:T1}-\eqref{eq:T2}. The transformed Moyal product, `$\sprT$' must satisfy the condition, \cite{Blaszak Domanski},
\begin{align}
(f \spr g) \circ T = (f \circ T) \sprT (g \circ T),~~~  f,g \in S(W) 
\end{align}
The transformed Moyal product is then given by,
\begin{align} \label{eq:New MP}
f \sprT g = (f\circ T) \, \exp \left( \frac{i\,\pp}{2} \left( \overleftarrow{D_{\KGnv}} \overrightarrow{D_{\KGnm}}\, - \overleftarrow{D_{\KGnm}} \, \overrightarrow{D_{\KGnv}} \right) + \frac{i\,\pp^2}{2} \left( \overleftarrow{D_{\Gnv}} \,\overrightarrow{D_{\Gnm}} - \overleftarrow{D_{\Gnm}} \,\overrightarrow{D_{\Gnv}} \right) \right)\, (g \circ T), 
\end{align}
with $(D_{\KGnv}, D_{\KGnm}, D_{\Gnv}, D_{\Gnm})$ being derivations transformed by $T$ according to,
\begin{align}
(\partial_{\KGv} f) \circ T &= \left(-\pp\,\Gnv\,\KGnv\,\partial_{\Gnv} + \pp\,\Gnm\,\KGnv\,\partial_{\Gnm} + \Gnv\,\Gnm\,\partial_{\KGnm} \right) (f\circ T) =: D_{\KGnm}(f\circ T) \label{eq:New Der phi}\\
(\partial_{\KGm} f) \circ T &= \left(-\pp\, \frac{\KGnm}{\Gnv\,(\Gnm)^2}\,\partial_{\Gnv} + \pp\,\frac{\KGnm}{\Gnv^2\,\Gnm}\,\partial_{\Gnm} + \frac{1}{\Gnv\,\Gnm}\,\partial_{\KGnv} \right)(f\circ T) =: D_{\KGnv}(f\circ T) \label{eq:New Der pi} \\
(\partial_{\Gv} f) \circ T &= \exp \left(-\pp\, \frac{\KGnv\,\KGnm}{\Gnv\,\Gnm} \right) \left( \left( 1+ \pp \,\frac{\KGnv\,\KGnm}{\Gnv\, \Gnm}  \right)\,\partial_{\Gnv} - \pp \frac{\KGnv\,\KGnm}{\Gnv^2}\, \partial_{\Gnm} - \frac{\KGnv}{\Gnv}\,\partial_{\KGnv} + \frac{\KGnm}{\Gnv}\, \partial_{\KGnm} \right) (f\circ T) \nonumber \\
&=: D_{\Gnv}(f\circ T) \label{eq:New Der a}\\
(\partial_{\Gm} f) \circ T &= \exp \left( \pp\, \frac{\KGnv\,\KGnm}{\Gnv\,\Gnm}\right) \left(  \pp\, \frac{\KGnv\,\KGnm}{(\Gnm)^2}\,\partial_{\Gnv} + \left(1 -\pp\, \frac{\KGnv\,\KGnm}{\Gnv\,\Gnm}\right) \,\partial_{\Gnm} - \frac{\KGnv}{\Gnm}\,\partial_{\KGnv} + \frac{\KGnm}{\Gnm}\, \partial_{\KGnm} \right) (f \circ T) \nonumber \\
&=: D_{\Gnm}(f\circ T) \label{eq:New Der pa}
\end{align}
The new Moyal product, \eqref{eq:New MP}, is in fact a well defined star product. It is straightforward to check that the first order of the exponential in \eqref{eq:New MP} yields simply the Poisson brackets with respect to the new variables $(\Gnv, \Gnm, \KGnv, \KGnm)$. For higher orders,  new contributions with respect to the transformed variables appear compared to the simple formula in \eqref{eq:Moyal Product} for the original variables. These must be taken into account in the transformed scheme.   \\

The canonical transformation of the previous section yields for the Hamilton symbol from \eqref{eq:Hamilton Symbol} in the new variables,
\begin{align}
\qf{h} = \irm{\tilde{E}}{hom} \cdot \Uf{\Hs{\te{p}}} + \frac{\exp (-\pp\, \frac{\KGnv\,\KGnm}{\Gnv\,\Gnm} )}{b} \sum_{\vec{m} \in \mathbbm{k}} \im{\tilde{\omega}}{\te{MS},\vec{m}} \,\adMSt{\vec{m}}\,\aMSt{\vec{m}}  + \frac{\exp \left(-\pp\, \frac{\KGnv\,\KGnm}{\Gnv\,\Gnm} \right)}{6\,b}\sum_{\vec{M} \in \mathbbm{K}} \im{\tilde{\omega}}{\text{T},\vec{m}} \;\bdTt{\vec{M}}{}\,\bTt{\vec{M}}{} 
\end{align}
The transformed Mukhanov-Sasaki and tensor frequencies are given by, 
\begin{align}
\im{\tilde{\omega}}{\te{MS},\vec{m}} := \pp^2 \sqrt{\hbox{$\vec{m}$}^{\,2} + \mMSt^2\,},~~~~  \im{\tilde{\omega}}{\text{T},\vec{m}}:= \pp^2 \sqrt{18\,\hbox{$\vec{m}$}^{\,2} + 6\,(\pp\, \mTt)^2\,},
\end{align}
with the respective masses defined according to \eqref{eq:mMS} and \eqref{eq:mT}. Note that the structure of the Fock space has remained unchanged. The annihilation and creation operators for the Mukhanov-Sasaki and the tensor system keep their dependence on the homogeneous phase space variables, however, transformed according to the canonical transformation.  Let $(\tilde{q},\tilde{p})$ be a short hand notation for the new homogeneous phase space variables $(\KGnv, \KGnm, \Gnv, \Gnm)$, and let $\tilde{x} \in (\tilde{q},\tilde{p})$. As for the non-transformed model, we can introduce connections $\qf{\mathcal{A}}^{\tilde{x}}$ for every new phase space variable $\tilde{x}$ in order to describe the derivatives of the Fock base states. Then, we write the derivatives of the base state $\eft{\nD}$ with respect to $\tilde{x}$ as,
\begin{equation} \label{eq:Def Connection}
\frac{\partial \, \eft{\nD}}{\partial \tilde{x}} =: (\qf{\mathcal{A}}^{\scriptscriptstyle{\tilde{x}}})\,\eft{\nD} =: \sum_k (\mathcal{A}^{\scriptscriptstyle{\tilde{x}}})_{\nD}^{~\,k}\, \eft{k},~~~~~   (\mathcal{A}^{\scriptscriptstyle{\tilde{x}}})_{\nD}^{~\,k} \in \mathbb{R},~\forall \lbrace \nD, k \rbrace,
\end{equation}
Analogously to the previous case, these connections connect states which are shifted by plus or minus two excitation numbers for every wave vector $\vec{m}$ seperately, namely,
\begin{align}
(\mathcal{A}^{\scriptscriptstyle{\tilde{x}}})_{\nD}^{~\,k} &= \sum_{\vec{m} \in \mathbbm{k}} \left( \delta_{\nD}^{~\lbrace ..., \nMS{m}-2,...\rbrace}\, \aonet{\nMS{m}}{\Lms,\tilde{x}} +  \delta_{\nD}^{~\lbrace ..., \nMS{m}+2,...\rbrace}\, \atwot{\nMS{m}}{\Lms,\tilde{x}} \right)  \\
& + \sum_{\vec{M} \in \mathbbm{K}} \left(  \delta_{\nD}^{~\lbrace ..., \nT{M}-2,...\rbrace}\,\aonet{\nT{M}}{\Lt,\tilde{x}}+ \delta_{\nD}^{}\atwot{\nT{M}}{\Lms,\tilde{x}}   \delta_{\nD}^{~\lbrace ..., \nT{M}+2,...\rbrace}\,\right) \nonumber 
\end{align}
where $\lbrace ..., \nMS{m}-2,...\rbrace$ denotes the set of excitation numbers which is shifted compared to the ``relevant'' excitation number $\nD$ by $-2$ in the single excitation number for the wave vector $\vec{m}$ in the Mukhanov-Sasaki state. Accordingly, for the other shifted excitation numbers. The $\delta$'s are thus Kronecker deltas. The $\tilde{\alpha}$-factors are explicitely given by,
\begin{align}
\aonet{\nMS{m}}{\Lms,x} &:= - \tilde{f}_{\Lms,\vec{m}}^{\,\tilde{x}} \frac{\sqrt{(\nMS{m}- 1)\cdot \nMS{m}}}{2}, \\
\atwot{\nMS{m}}{\Lms,x} &:= \tilde{f}_{\Lms,\vec{m}}^{\,\tilde{x}} \frac{\sqrt{(\nMS{m}+ 1)\cdot (\nMS{m} +2)}}{2}, \\
\aonet{\nT{M}}{\Lt,x} &:= - \tilde{f}_{\Lt,\vec{m}}^{\,\tilde{x}} \frac{\sqrt{(\nT{M}- 1)\cdot \nT{M}}}{2}, \\
\atwot{\nT{M}}{\Lt,x} &:= \tilde{f}_{\Lt,\vec{m}}^{\,\tilde{x}} \frac{\sqrt{(\nT{M}+ 1)\cdot (\nT{M} +2)}}{2},
\end{align}
where the functions $\tilde{f}$ derive from the frequencies,
\begin{equation}
\tilde{f}_{\Lms,\vec{m}}^{\,\tilde{x}} = - \frac{\partial_{\tilde{x}} \mMSt^2}{4\, \left(\hbox{$\vec{m}$}^{\,2} + \mMSt^2 \right)}, ~~~~ \tilde{f}_{\Lt,\vec{M}}^{\,\tilde{x}} = - \frac{3\,\pp^2\,\partial_{\tilde{x}} \mTt^2}{2\, \left( 18\,\hbox{$\vec{m}$}^{\,2} + 6\,(\pp\, \mTt)^2 \right)}.
\end{equation}
In order to derive the effective Hamiltonian within some Fock subspace with quantum number(s), $\nD$, we can come back to the equation of the previous model in \eqref{eq:heff 0}, \eqref{eq:heff1} and \eqref{eq:heff 2}. The results for the zeroth and first order effective Hamiltonian translate without any modifications to the transformed model: At zeroth order, the effective Hamilton symbol is given by the Born-Oppenheimer approximation,
{\footnotesize
\begin{align} \label{eq:heff t 0}
\heff{,0,\te{r}}{} =  \sum_{\delta} \eftzero{\nD}\scpr{\eftzero{\nD}}{\cdot} \left(\irm{\tilde{E}}{hom} + \frac{\exp \left(-\pp\, \frac{\KGnv\,\KGnm}{\Gnv\,\Gnm} \right)}{b} \left( \sum_{\vec{m} \in \mathbbm{k}} \im{\tilde{\omega}}{\te{MS},\vec{m}}\, \im{n}{\te{MS},\vec{m},d} + \frac{1}{6}\,\sum_{\vec{M} \in \mathbbm{K}} \im{\tilde{\omega}}{\te{T},\vec{M}}\, \im{n}{\te{T},\vec{M},d'} \right) \right).
\end{align}} 

Due to the conservation of the Poisson bracket, the first order effective Hamiltonian, $\heff{,1,\te{r}}{}$ vanishes, similar to the previous model. \\

The second order effective Hamiltonian must be determined according to \eqref{eq:heff2 app}, but with Poisson brackets with respect to the new variables, and the former star product `$\spr$' replaced by the star product, `$\sprT$'. For the contributions to the effective Hamiltonian which do not involve the second order Moyal product, i.e., the first four and the last terms in \eqref{eq:heff2 app}, we can simply take the preceding results, perform the canonical transformation for the involved symbols and replace the derivatives with respect to $(\Gv, \Gm, \KGv, \KGm)$ by derivatives with respect to $(\Gnv, \Gnm, \KGnv, \KGnm)$. Namely, we use \eqref{eq:heff2 explicit} and perform the above transformations. A similar reasoning regarding the orders in $\pp$ of the respective contributions, reveals that the resulting effective Hamilton constraint for the given perturbative truncation contains only a finite number of terms. In particular, the only relevant terms are,
{\small
\begin{align}
&\frac{i}{2}\, \lbrace \pli{\te{R}}\,\us{1}{}, \qf{h} + \im{E}{n} \rbrace \,\us{0}{\ast} \,\pli{\te{R}} =  \sum_{\delta} \eftzero{\nD}\scpr{\eftzero{\nD}}{\cdot} \nonumber \\
& \cdot \left\lbrace \sum_{\vec{m} \in \mathbbm{k}} \left(-\frac{\Gnv\,\exp \left(\pp\, \frac{\KGnv\,\KGnm}{\Gnv\,\Gnm}\right)}{64}\,\frac{2\,\nMS{m} +1}{\pp^2 \left(\hbox{$\vec{m}$}^{\,2} + \mMSt^2\,\right)^{5/2} }\cdot \left((\partial_{\KGnm} \mMS^2)\,(\partial_{\KGnv} \irm{\tilde{E}}{hom}) - (\partial_{\KGnv} \mMSt^2)\,(\partial_{\KGnm} \irm{\tilde{E}}{hom}) \right)^2 \right) \right. \nonumber \\
& ~+ \left. \sum_{\vec{M} \in \mathbbm{K}} \left(-\frac{27\,\Gnv\,\exp \left(\pp\, \frac{\KGnv\,\KGnm}{\Gnv\,\Gnm}\right)}{8}\,\frac{2\,\nT{M} +1}{\pp^2 \left(18\,\hbox{$\vec{m}$}^{\,2} + 6\,(\pp \mTt)^2\,\right)^{5/2} } \cdot \left((\partial_{\KGnm} (\pp \mTt)^2)\,(\partial_{\KGnv} \irm{\tilde{E}}{hom}) \right)^2 \right)  \right\rbrace. \label{eq:heff2 T}
\end{align}}
The contributions of the second order effective Hamiltonian, arising from the second order Moyal product in \eqref{eq:heff2 app}, result from applying the transformed Moyal product, `$\sprT$', from \eqref{eq:New MP}, together with, \eqref{eq:New Der phi} -- \eqref{eq:New Der pa}. The computations are straightforward but lengthy. However, all possible terms are of higher order in $\pp$ than the considered truncation: First, the double Moyal product enters at least with powers $\pp^2$, also with respect to the new variables, as it can directly be seen from the transformation formulae. Second, we consider the Moyal product of $\us{0}{}$ and $(\qf{h}- \im{E}{n})$, where the second term and derivatives thereof, enter at least with a factor $\pp^2$, too. Since the connections and their derivatives never enter with negative powers in $\pp$, all contributions are of higher than second order in $\pp$. Indeed, the only relevant contributions for the effective Hamiltonian at second order stem from \eqref{eq:heff2 T}.  \\

We present the result for the effective Hamiltonian with respect to the transformed variables $(\Gnv, q, \KGnv, u)$, i.e., without the $\pp$-scaling for the momentum variables. It consists of the zeroth order contribution, \eqref{eq:heff t 0}, and the second order contribution, \eqref{eq:heff2 T}. Expressing the latter explicitly as a function of the transformed variables, we obtain, 
{\small 
\begin{align}
\heff{,(2),\te{r}}{} =&  \sum_{\delta} \eftzero{\nD}\scpr{\eftzero{\nD}}{\cdot} \nonumber \\
& \cdot \left\lbrace \, \frac{L^3}{2} \frac{\pp^2\,q^2}{\Gnv^2} \exp \left(- \frac{\KGnv\,u}{\Gnv\,q} \right)\left( \KGnv^2 - \frac{1}{6} \right)  \right. \label{eq:hefft 1}\\
& ~~~~  + \frac{\exp \left(- \frac{\KGnv\,u}{\Gnv\,q} \right)}{b} \left( \sum_{\vec{m} \in \mathbbm{k}} \sqrt{\hbox{$\vec{m}$}^{\,2} + \mMSt^2}\, \im{n}{\te{MS},\vec{m},d} + \frac{1}{6}\,\sum_{\vec{M} \in \mathbbm{K}} \sqrt{18\,\hbox{$\vec{m}$}^{\,2} + 6\,(\pp\,\mTt)^2\,}\, \im{n}{\te{T},\vec{M},d'} \right) \label{eq:hefft 2} \\
& ~~~~ + \left. \sum_{\vec{m} \in \mathbbm{k}} \frac{2\,\im{n}{\te{MS},\vec{m},d} +1}{(\hbox{$\vec{m}$}^{\,2} + \mMSt^2)^{5/2}} \cdot \irm{\tilde{h}}{MS} +  \sum_{\vec{M} \in \mathbbm{K}} \frac{2\,\im{n}{\te{T},\vec{M},d'} +1}{(18\,\hbox{$\vec{m}$}^{\,2} + 6\,(\pp\,\mTt)^2)^{5/2}} \cdot \irm{\tilde{h}}{T} \right\rbrace \label{eq:hefft 3}
\end{align}}

where we introduced the Hamiltonian backreaction functions,
{\footnotesize 
\begin{align}
\irm{\tilde{h}}{MS} &= \exp \left(-\,13\,\frac{\KGnv\,u}{\Gnv\,q} \right) \left( -\frac{81\, \pp^{12}\, q^6\, \KGnv^4 \left(2\, c_+ \,\KGnv^2+c_- \left(-8\, c_+ +2\, \KGnv^2+1\right)+c_+ + 4\, \KGnv^4-2\, \KGnv^2\right){}^2}{64\, \Gnv^7} \right),\\
\irm{\tilde{h}}{T} &= - \exp \left(-\,13\, \frac{\KGnv\,u}{\Gnv\,q} \right)\cdot \frac{3\,\pp^{12}\,q^6 \KGnv^4}{2\, \Gnv^7}.
\end{align}}

In the model with canonical variables $(\Gnv,q,\KGnv,u)$, we obtain the standard Born-Oppenheimer Hamiltonian, 
\eqref{eq:hefft 1}, \eqref{eq:hefft 2} as it is the case for the model in the original variables. However, the backreaction terms from the second space adiabatic order now include both Mukhanov-Sasaki \emph{and} tensor contributions \eqref{eq:hefft 3} which are again of the {\it Casimir type}. The reason for this to happen is that the restriction to the positive mass region is accomplished by a symplectic embedding rather than a symplectomorphism which in particular changes the entire topology of the slow phase space. Thus, the quantum theories cannot be unitarily equivalent. Note that even if the phase spaces were the same and the transformation was strictly canonical, the Moyal products do not simply get rewritten in terms of the new variables unless the canonical transformation is of a restricted type called ``gauge equivalent'' in \cite{Blaszak Domanski}.

\section{Conclusion and Outlook}

In this paper we applied space adiabatic perturbation theory to the hybrid approach to quantum cosmological perturbation theory and demonstrated that the challenges due to the quantum field theory nature of the problem for which space adiabatic perturbation theory was not designed can be faced squarely. What we now have at our disposal is a machinery to compute the corrections 
from every energy band of the inhomogeneous Fock space to the effective homogeneous Hamiltonian in principle to arbitrary adiabatic order. In this paper, we carried out this programme to second adiabatic order. While the computations are rather tedious already to this order, it is clear how to proceed to arbitrary order, the scheme is similar in nature to standard textbook quantum mechanical perturbation theory of pure point spectra.

The treatment of the backreaction problem beyond the semiclassical regime, which we believe to be very important especially in the Planck era of the universe where we expect the semiclassical approximation to be poor, has revealed many new interesting challenges
including: Hilbert-Schmidt conditions that require the mixture of homogeneous and inhomogeneous degrees of freedom in which to formulate the quantum field theory, Weyl-Moyal calculus to feed in the quantum nature of the background into the Fock space construction of the inhomogeneous sector, tachyons and their avoidance and adiabatic backreactions of the Casimir type. Note that, in contrast to the plain Casimir term in QED, this contribution converges. It is not clear from the current status of the calculations that this also happens at higher adiabatic order in which case presumably also non-trivial renormalisations would be required.    

What is left to do is to evaluate the phenomenological consequences of this computation and to compare to similar ones in the literature which mostly focusses on the vacuum energy band. From the explicit expression of our end result it is evident that the corrections to the effective homogeneous contribution of the Hamiltonian constraint
that one unambiguously obtains from the space adiabatic 
treatment of the backreactions are, unsurprisingly, of a rather new type not previously encountered in more semiclassical treatments \cite{8a,8b,8c} and it will be 
interesting to see how these terms affect the previous analysis of the effective dynamics.
 
In particular, in view of \cite{12} we are in the position to use the standard Schr\"odinger representation to quantise the effective homogeneous Hamiltonian which in view of the Weyl quantisation techniques used by space adiabatic perturbation theory is more natural than the Loop Quantum Cosmology representation and not plagued by discretisation ambiguities as one can directly quantise position and momentum operators rather than Weyl element approximants. The results of this investigation are reserved for a future publication.\\  
\\
\\
{\bf Acknowledgements}\\
\\
S.S. thanks the
Heinrich-B\"oll Stiftung for financial and intellectual support and the
German National Merit Foundation for intellectual support.

\begin{appendix}

\section{Selected Details of Quantum Cosmological Space Adiabatic Perturbation Theory to Second Adiabatic Order}
We provide the explicit expressions for sections \ref{sec:Moyal Proj} to \ref{sec:Eff Ham}. \\

The space adiabatic scheme employs the Moyal or star product ``$\spr$'' for the homogeneous degrees of freedom $(q,p) \in \Gamma$. Let therefore $A^{(q,p)}, B^{(q,p)} \in S(\delta; \Gamma, \mathcal{L}(\Hs{\te{p}})$ be two semiclassical symbols with respect to the perturbation parameter $\delta$ and with formal series expansion, $A \asymp \sum_j \delta^j A_j$, $B \asymp \sum_l \delta^l B_l$. In general, the Moyal product for two symbols $A,B$ provides a formal power series with respect to $\delta$. We refer to \cite{Teufel}, in particular to Appendix A therein, for a self-contained review of pseudodifferential calculus. We note that for our model, two perturbation parameters enter the game; the perturbation parameter for the homogeneous scalar field variables $(\KGv,\KGm)$ is $\pp$, while the perturbation parameter for the homogeneous gravitational variables $(\Gv,\Gm)$ is $\pp^2$. $(q,p)$ is then a short hand notation for the whole set of homogeneous phase space variables. The star product is consequently defined as,
\begin{align} \label{eq:Moyal Product}
\left(A \spr B\right)^{(q,p)} \asymp&~ A\,\exp \left( \frac{i \,\pp}{2} \left(\overleftarrow{\partial_{\KGv}} \,\overrightarrow{\partial_{\KGm}} - \overleftarrow{\partial_{\KGm}} \,\overrightarrow{\partial_{\KGv}} \right) + \frac{i\, \pp^2}{2} \left(\overleftarrow{\partial_{\Gv}} \,\overrightarrow{\partial_{\Gm}} - \overleftarrow{\partial_{\Gm}} \, \overrightarrow{\partial_{\Gv}} \right)  \right) B.
%& = \, \sum_{k=0}^{\infty} \pp^k (2i)^{-k} \sum_{|\alpha|+|\beta|+j+l =k} \frac{(-1)^{|\alpha|}}{|\alpha|! |\beta|!} \left((\partial^{\alpha}_{\KGm} \partial^{\beta}_{\KGv} A_j) (\partial^{\alpha}_{\KGv} \partial^{\beta}_{\KGm} B_l) \right)^{(q,p)} \nonumber \\
%& + \sum_{k=0}^{\infty} \pp^{2k} (2i)^{-k} \sum_{|\alpha|+|\beta|+j+l =k} \frac{(-1)^{|\alpha|}}{|\alpha|! |\beta|!} \left((\partial^{\alpha}_{\Gm} \partial^{\beta}_{\Gv} A_j) (\partial^{\alpha}_{\Gv} \partial^{\beta}_{\Gm} B_l) \right)^{(q,p)}. \label{eq:Moyal Product}
\end{align}
As it turns out, the Moyal product with respect to the gravitational degrees of freedom does not yield any contributions to the computations for the considered order of the perturbation theory since it involves the higher-order perturbation parameter. Also note that the Moyal product involves the Poisson bracket with a factor $i/2$ at first perturbative order. Thus, we employ $\im{\lbrace \cdot, \cdot \rbrace}{(\KGv,\KGm)} \!=: \lbrace \cdot, \cdot \rbrace$ as the Poisson bracket with respect to the homogeneous scalar field variables. \\

We start with the construction of the Moyal projector at first order, $\pli{1}$. The symbol can be split into diagonal and off-diagonal parts with respect to the basis choice in \eqref{eq:EVP},   $\pli{1} =: \pli{1}^{\te{D}} + \pli{1}^{\te{OD}}$. The diagonal part is simply defined as, $\pli{1}^{\te{D}} = \pli{0}\,\pli{1} \,\pli{0} + (\Uf{\Hs{\te{p}}}\! \!-\pli{0})\,\pli{1}(\Uf{\Hs{\te{p}}}\! \!-\pli{0})$. \\
As notified in \ref{sec:Moyal Proj}, the only relevant contributions for the construction of the effective Hamiltonian up to second order, are the off-diagonal ones, $\pli{1}^{\te{OD}}$. The defining equations for $\pli{1}^{\te{OD}}$ result from condition $3)$ in \ref{sec:Moyal Proj} and by means of the Moyal product \eqref{eq:Moyal Product}. We obtain defining equations for every considered order in $\pp$. They result by carefully executing the Moyal product and collecting the terms at the considered order in $\pp$. \\
The zeroth order equation resulting from condition $3)$ is satisfied by construction, $\pli{0} \,\qf{h} - \qf{h}\, \pli{0} = 0$, where the product is the operator product in $\mathcal{L}(\Hs{\te{p}})$. Note that in this scheme, $\qf{h} = \im{\qf{h}}{0}$, i.e., $\qf{h}$ is already the zeroth order Hamiltonian symbol. At first order, we do not only have $\pli{1}$ entering the equations, but also the Poisson bracket with respect to $(\KGv, \KGm)$ according to \eqref{eq:Moyal Product}. Accordingly, the defining equation for $\pli{1}^{\te{OD}}$ is straightforwardly given by,
\begin{equation} \label{eq:Pi1 Def Equ}
\left[\,\qf{h}, \pli{1}^{\te{OD}} \right] = - \left[\qf{h}, \pli{1}^{\te{D}} \right] - \frac{i}{2}\, \lbrace \qf{h},\pli{0} \rbrace + \frac{i}{2}\, \lbrace \pli{0}, \qf{h} \rbrace.
\end{equation}
The explicit expression for $\pli{1}^{\te{OD}}$ follows from first multiplying by $\pli{0}$ and by $(\Uf{\Hs{\te{p}}}-\pli{0})$ from the left and the right respectively, and repeating this operation with the operators exchanged. Since we consider only \emph{one} relevant energy value $\im{E}{n}$, the factor $\im{E}{n}$ can then be drawn from one side to another if necessary. The first contribution in \eqref{eq:Pi1 Def Equ} vanishes because of the multiplication operations, while the latter terms remain. By summing the two resulting contributions, it is straightforward to obtain the final result in \eqref{eq:Pi1}, i.e.,
\begin{align}
\im{\qf{\pi}}{1}^{\scriptscriptstyle{\te{OD}}} =& - \frac{i}{2} \left( \qf{h}- \!\im{E}{n} \right)^{-1} \!\left( \Uf{\Hf}\! \!- \im{\qf{\pi}}{0} \right) \!\lbrace \qf{h} + \im{E}{n}, \im{\qf{\pi}}{0} \rbrace \,\im{\qf{\pi}}{0} - \frac{i}{2} \, \im{\qf{\pi}}{0} \im{\lbrace \qf{\pi}}{0}, \qf{h} + \im{E}{n} \rbrace \left( \qf{h}-\! \im{E}{n} \right)^{-1}\! \left( \Uf{\Hf} \! \!- \im{\qf{\pi}}{0} \right). \nonumber
\end{align}
Recall that the Poisson bracket involves the derivatives with respect to $\KGv$ and $\KGm$. As a showcase example, the Poisson bracket in the first term evaluates to,
\begin{align} \label{eq:Evaluation Poisson Bracket}
&\lbrace \qf{h} + \im{E}{n}, \im{\qf{\pi}}{0} \rbrace \,\im{\qf{\pi}}{0}  = \left( \frac{\partial (\qf{h} + \im{E}{n})}{\partial \KGv} \frac{\partial \pli{0}}{\partial \KGm} -\frac{\partial (\qf{h} + \im{E}{n})}{\partial \KGm} \frac{\partial \pli{0}}{\partial \KGv}  \right) \im{\qf{\pi}}{0} \\
~ & ~ = \sum_m \left( \frac{\partial (\im{E}{m} \!+ \!\im{E}{n})}{\partial \KGv} \ef{m} \scpr{\ef{m}}{\cdot} + \im{E}{m} \left( \frac{\partial \ef{m}}{\partial \KGv} \scpr{\ef{m}}{\cdot} + \ef{m} \scpr{ \frac{\partial \ef{m}}{\partial \KGv}}{\cdot} \right) \right) \frac{\partial \ef{\nD}}{\partial \KGm} \scpr{\ef{\nD}}{\cdot} \nonumber \\
& ~~~  - \sum_m \left( \frac{\partial (\im{E}{m} \!+ \!\im{E}{n})}{\partial \KGm} \ef{m} \scpr{\ef{m}}{\cdot} + \im{E}{m} \left( \frac{\partial \ef{m}}{\partial \KGm} \scpr{\ef{m}}{\cdot} + \ef{m} \scpr{ \frac{\partial \ef{m}}{\partial \KGm}}{\cdot} \right) \right) \frac{\partial \ef{\nD}}{\partial \KGv} \scpr{\ef{\nD}}{\cdot} \nonumber 
\end{align}
The derivatives of the basis states $\ef{m}$ with respect to $\KGv$ and $\KGm$ involve the connection as defined in \eqref{eq:Def Connection}, in particular, $\partial_x \e{m} = (\qf{\mathcal{A}}^{\scriptscriptstyle{x}})\e{m} = \sum_j (\!\mathcal{A}^{\scriptscriptstyle{x}}\!)_{m}^{~~j} \e{j}$. Introducing the connections into \eqref{eq:Evaluation Poisson Bracket} and using that both, the wave functions $\e{m}$ and the entries of the connections $(\!\mathcal{A}^{\scriptscriptstyle{x}}\!)_{m}^{~~j}$, are real-valued, we obtain,
\begin{align}
\pli{1}^{\te{OD}} =& \,- \frac{i}{2} \sum_{\delta} \sum_{m \neq \nD} \left( e_{\nD}^{\qpb} \scpr{e_m^{\qpb}}{\cdot} - e_{m}^{\qpb} \scpr{e_{\nD}^{\qpb}}{\cdot}\right) \cdot \frac{1}{\im{E}{m}\!-\!\im{E}{n}}\\
& \cdot \left\lbrace  (\!\mathcal{A}^{\scriptscriptstyle{\KGv}}\!)_{\nD}^{~m} \frac{\partial (\im{E}{m}\! + \!\im{E}{n})}{\partial \KGm} -  (\!\mathcal{A}^{\scriptscriptstyle{\KGm}}\!)_{\nD}^{~m} \frac{\partial (\im{E}{m}\! + \!\im{E}{n})}{\partial \KGv} + \sum_j \left( (\!\mathcal{A}^{\scriptscriptstyle{\KGv}}\!)_{\nD}^{~j} (\!\mathcal{A}^{\scriptscriptstyle{\KGm}}\!)_{j}^{~m} - (\!\mathcal{A}^{\scriptscriptstyle{\KGm}}\!)_{\nD}^{~j} (\!\mathcal{A}^{\scriptscriptstyle{\KGv}}\!)_{j}^{~m} \right) \im{E}{j}\right\rbrace, \nonumber
\end{align}
and we recall that $\delta$ is the degeneracy index, $\nD$ is the set of quantum numbers of the perturbation system for the relevant subspace. The sum over $m$ includes in principle \emph{all} possible sets of excitation numbers for the Mukhanov-Sasaki and the tensor field system. However, the entries of the connections, $(\!\mathcal{A}^{\scriptscriptstyle{\KGv}}\!)_{\nD}^{~m}$ and $(\!\mathcal{A}^{\scriptscriptstyle{\KGm}}\!)_{\nD}^{~m}$, are only non-vanishing for $m$'s which differ by plus or minus two excitations for one single wave vector compared to the set of excitation numbers $\nD$, as it can be seen from the explicit evaluation of the connections in, \eqref{eq:Derivative States}. \\

The next step consists in computing the Moyal unitary at first order, $\us{1}{}$. In close analogy to the construction scheme in \cite{Teufel}, we introduce a hermitian part of $\us{1}{}$, namely $\im{\qf{a}}{1}\,\us{0}{}$ and an anti-hermitian part, $\im{\qf{b}}{1}\,\us{0}{}$, such that, $\im{\qf{a}}{1}^{\ast} = \im{\qf{a}}{1}$ and $\im{\qf{b}}{1}^{\ast} = -\im{\qf{b}}{1}$, and in total $\us{(1)}{} =: \us{0}{} + \pp \left(\im{\qf{a}}{1} + \im{\qf{b}}{1} \right) \,\us{0}{}$. Then, the conditions $2)$ or $3)$ in section \ref{sec:Moyal Unitary} define the hermitian part, $\im{\qf{a}}{1}$, while condition $1)$ defines the anti-hermitian part, $\im{\qf{b}}{1}$. \\
For the hermitian part, the procedure requires to evaluate the star product for $\us{(1)}{}$ together with $\us{(1)}{}$. At zeroth order, condition $2)$ is trivially satisfied: $\us{0}{} \cdot \us{0}{\ast} - \Uf{\Hs{\te{p}}} = 0$. At first order, we simply get,
\begin{equation}
\im{\qf{a}}{1} = -\frac{i}{4}\,\lbrace \us{0}{}, \us{0}{\ast} \rbrace.
\end{equation}
The evaluation of this symbol by means of the connections immediately reveals that $\im{\qf{a}}{1} $ vanishes, again because the connections and the states are real-valued. This can be seen directly by evaluating the Poisson bracket as in \eqref{eq:Evaluation Poisson Bracket}. \\
For the anti-hermitian part of $\us{1}{}$, it is necessary to evaluate the double star product in condition $1)$ of section \ref{sec:Moyal Unitary}. A straightforward application of the Moyal product, \eqref{eq:Moyal Product}, and the comparison of terms at first order in $\pp$, shows that $\im{\qf{b}}{1}$ must satisfy,
\begin{equation}
\left[\, \im{\qf{b}}{1}, \pli{\te{R}}\, \right] = - \left( \im{\qf{a}}{1}\, \pli{\te{R}} + \pli{\te{R}}\, \im{\qf{a}}{1} + \us{0}{}\,\pli{1}\,\us{0}{\ast} + \frac{i}{2}\, \us{0}{} \lbrace \pli{0}, \us{0}{\ast} \rbrace + \frac{i}{2}\,\lbrace \us{0}{}, \pli{0}\,\us{0}{\ast} \rbrace 
 \right).
\end{equation}
A solution to this is simply, $\im{\qf{b}}{1} = - \left[ \,\pli{\te{R}}, \,\left[\, \im{\qf{b}}{1}, \pli{\te{R}}\, \right] \right]$. Since $\im{\qf{a}}{1}$ vanishes and because $\pli{0}\,\us{0}{\ast} = \us{0}{\ast}\,\pli{\te{R}}$, it follows that, 
\begin{align} \label{eq:u1 Appendix}
\us{1}{} &= \im{\qf{b}}{1}\,\us{0}{} = \left[\irm{\qf{\pi}}{R}, \im{\qf{u}}{0}\,\im{\qf{\pi}}{1}^{\te{OD}}\, \im{\qf{u}}{0}^{\ast} \right]\,\im{\qf{u}}{0} \\
&= - \frac{i}{2}\, \pli{\te{R}}\, \us{0}{} \lbrace \pli{0}, \qf{h} + \im{E}{n} \rbrace (\qf{h}-\im{E}{n})^{-1} (\Uf{\Hs{\te{p}}}\!\!-\pli{0}) + \frac{i}{2}\, \us{0}{} (\qf{h} - \im{E}{n})^{-1} (\Uf{\Hs{\te{p}}}\!\!-\pli{0}) \lbrace \qf{h} + \im{E}{n}, \pli{0} \rbrace\, \pli{0} \nonumber
\end{align}
In terms of the connections, $\us{1}{}$ is given by,
\begin{align}
\us{1}{} =& \,- \frac{i}{2} \sum_{\delta} \sum_{m \neq \nD} \left( e_{\nD}^{\qpbzero} \scpr{e_m^{\qpb}}{\cdot} + e_{m}^{\qpbzero} \scpr{e_{\nD}^{\qpb}}{\cdot}\right) \cdot \frac{1}{\im{E}{m}\!-\!\im{E}{n}}\\
& \cdot \left\lbrace  (\!\mathcal{A}^{\scriptscriptstyle{\KGv}}\!)_{\nD}^{~m} \frac{\partial (\im{E}{m}\! + \!\im{E}{n})}{\partial \KGm} -  (\!\mathcal{A}^{\scriptscriptstyle{\KGm}}\!)_{\nD}^{~m} \frac{\partial (\im{E}{m}\! + \!\im{E}{n})}{\partial \KGv} + \sum_j \left( (\!\mathcal{A}^{\scriptscriptstyle{\KGv}}\!)_{\nD}^{~j} (\!\mathcal{A}^{\scriptscriptstyle{\KGm}}\!)_{j}^{~m} - (\!\mathcal{A}^{\scriptscriptstyle{\KGm}}\!)_{\nD}^{~j} (\!\mathcal{A}^{\scriptscriptstyle{\KGv}}\!)_{j}^{~m} \right) \im{E}{j}\right\rbrace. \nonumber
\end{align}
Finally, we turn to the evaluation of the effective Hamiltonian $\heff{,(2),\te{r}}{}$. As already stated, the zeroth order corresponds to the energy within the relevant subspace of $\im{E}{n}$ and does not require any further manipulations. \\
For the first order, we observe that the defining equation of $\heff{}{}$ in \eqref{eq:heff def} gives after star multiplying the Moyal unitary $\us{}{}$ from the right,
\begin{equation} \label{eq:heff1 reasoning}
\us{}{} \spr \qf{h} - \heff{,0}{} \spr \us{}{} \,=\, \pp\,\heff{,1}{} \spr \us{}{} + \mathcal{O}(\pp^2) \,=\, \pp\,\heff{,1}{} \us{0}{} + \mathcal{O}(\pp^2).
\end{equation}
Here, the second equality follows from putting all contributions of second order and higher in $\pp$ into $\mathcal{O}(\pp^2)$. It is then straightforward to evaluate the left hand side of the equation, recalling that $\qf{h}$ is already the zeroth order symbol while $\us{}{}$ has both, a zeroth and a first order contribution. Then, multiplication by $\us{0}{\ast}$ from the right yields the result in \eqref{eq:heff1}, namely,
\begin{equation} 
\heffscp{,1} = \left( -\, \heff{,0}{}\,\im{\qf{u}}{1}  + \im{\qf{u}}{1}\, \qf{h} + \frac{i}{2} \im{\lbrace \im{\qf{u}}{0}, \qf{h} \rbrace}{\phi} - \frac{i}{2}  \im{\lbrace \heff{,0}{} , \im{\qf{u}}{0} \rbrace}{\phi}\right) \im{\qf{u}}{0}^{\ast}. \nonumber
\end{equation}
Regarding the restriction to the relevant subspace, note that the multiplication from $\pli{\te{R}}$ from the left and from the right on $\heff{,1}{}$, gives,
\begin{align}
\heff{,1,\te{r}}{} &= \pli{\te{R}}\,\heff{,1}{} \,\pli{\te{R}} \nonumber \\
&= \pli{\te{R}} \,\us{1}{} (\qf{h}- \im{E}{n})\, \us{0}{\ast}\,\pli{\te{R}} + \frac{i}{2}\, \pli{\te{R}} \lbrace \us{0}{}, \qf{h} + \im{E}{n} \rbrace\, \us{0}{\ast}\, \pli{\te{R}} \nonumber \\
&= \frac{i}{2}\, \pli{\te{R}} \lbrace \us{0}{}, \qf{h} + \im{E}{n} \rbrace\, \us{0}{\ast}\, \pli{\te{R}} 
\end{align}
The first term in the second line vanishes trivially since the restriction of $\qf{h}$ onto $\pli{\te{R}}$ from the right is just $\im{E}{n}$, which cancels the other term. The second contribution vanishes due to symmetry. We show this by employing the representation by means of the connections,
\begin{align}
\frac{i}{2}\, \pli{\te{R}} \lbrace \us{0}{}, \qf{h} + \im{E}{n} \rbrace\, \us{0}{\ast}\, \pli{\te{R}}  &= \frac{i}{2}\,\sum_{\delta} \efzero{\nD} \scpr{\efzero{\nD}}{\cdot} \nonumber \\
&~~ \cdot \left\lbrace \sum_m \left( (\!\mathcal{A}^{\scriptscriptstyle{\KGv}}\!)_{\nD}^{~m} \frac{\partial(\im{E}{m} \!+ \!\im{E}{n})}{\partial \KGm} - (\!\mathcal{A}^{\scriptscriptstyle{\KGm}}\!)_{\nD}^{~m} \frac{\partial(\im{E}{m} \!+ \!\im{E}{n})}{\partial \KGv}  \right) \scpr{\ef{m}}{\ef{\nD}} \right. \nonumber \\
&~~~~~~  + \left. \sum_m \left( (\im{E}{n}\! -\! \im{E}{m})\cdot \left((\!\mathcal{A}^{\scriptscriptstyle{\KGv}}\!)_{\nD}^{~m} (\!\mathcal{A}^{\scriptscriptstyle{\KGm}}\!)_{\nD}^{~m}  - (\!\mathcal{A}^{\scriptscriptstyle{\KGm}}\!)_{\nD}^{~m} (\!\mathcal{A}^{\scriptscriptstyle{\KGv}}\!)_{\nD}^{~m} \right) \right) \right\rbrace \nonumber\\
& = 0. \label{eq:heff1 vanishes}
\end{align}
The second line vanishes because the connections $(\!\mathcal{A}^{\scriptscriptstyle{x}}\!)_{\nD}^{~m}$ have vanishing diagonal entries while the scalar product between $\e{m}$ and $\e{\nD}$ vanishes for $m\neq\nD$. The third line vanishes identically due to the second factor involving the connections. This is because the connection with respect to $\KGm$ only connects states which are shifted by Mukhanov-Sasaki mode excitations, and consequently only the Mukhanov-Sasaki mode shifts of $(\!\mathcal{A}^{\scriptscriptstyle{\KGv}}\!)_{\nD}^{~m}$ are relevant. In this case, however, $(\!\mathcal{A}^{\scriptscriptstyle{\KGv}}\!)_{\nD}^{~m}$ and $(\!\mathcal{A}^{\scriptscriptstyle{\KGm}}\!)_{\nD}^{~m}$ differ only by an overall factor and the connection part in this line vanishes trivially. \\

The second order contribution of the effective Hamiltonian derives from the same arguments as for $\heff{,1}{}$. In fact, we obtain in line with the reasoning in \eqref{eq:heff1 reasoning} at second perturbative order,
{\small 
\begin{align} \label{eq:heff2 app}
\heff{,2,\te{r}}{} =~ &\, \pli{\te{R}} \left( - \heff{,1}{}\,\us{1}{} + \frac{i}{2}\,\lbrace \us{1}{}, \qf{h} \rbrace - \frac{i}{2} \lbrace \im{E}{n}, \us{1}{} \rbrace - \frac{i}{2}\, \lbrace \heff{,1}{} ,\us{0}{} \rbrace + (\us{0}{} \spr \qf{h})_2 - (\im{E}{n} \spr \us{0}{})_2 \right)\,\us{0}{\ast}\,\pli{\te{R}} \nonumber \\
& + \frac{i}{2}\, \pli{\te{R}}\,\im{\lbrace \us{0}{}, \qf{h}\!+\!\im{E}{n} \rbrace}{(\Gv,\Gm)}\,\us{0}{}\,\pli{\te{R}}
\end{align}}

where the index ``$2$'' of the two latter contributions in the first line stands for the restriction of the Moyal product to the second order in $\pp$. We emphasize that $\pli{\te{R}}$ does not depend on the phase space variables $(q,p)$. Hence, its derivatives with respect to $(q,p)$ vanish. The second line involves the Poisson bracket with respect to the gravitational phase space variables $(\Gv,\Gm)$. We analyse the terms step by step. \\

The first contribution, $- \pli{\te{R}}\, \heff{,1}{}\,\us{1}{} \,\us{0}{\ast}\,\pli{\te{R}}$, vanishes identically. In order to see this, we only need to evaluate the left-hand part of it, namely,
\begin{align} 
\pli{\te{R}}\,\heff{,1}{} &= \pli{\te{R}}\,\us{1}{}\,(\qf{h}-\im{E}{n})\,\us{0}{\ast} + \frac{i}{2}\,\pli{\te{R}} \lbrace \us{0}{}, \qf{h} + \im{E}{n} \rbrace\,\us{0}{\ast} \nonumber \\
&= - \,\frac{i}{2}\, \pli{\te{R}}\,\us{0}{} \lbrace \pli{0}, \qf{h} + \im{E}{n} \rbrace (\Uf{\Hs{\te{p}}}\!\!-\pli{0})\, \us{0}{\ast} + \frac{i}{2}\,\pli{\te{R}} \lbrace \us{0}{}, \qf{h} + \im{E}{n} \rbrace\,\us{0}{\ast} \nonumber \\
&= \frac{i}{2}\, \pli{\te{R}} \lbrace \us{0}{}, \qf{h} + \im{E}{n} \rbrace\,\pli{0}\,\us{0}{\ast}, \label{eq:piRheff1}
\end{align}
where we induced the result of $\us{1}{}$ in \eqref{eq:u1 Appendix}. We have already shown in \eqref{eq:heff1 vanishes} that this operator vanishes identically. \\

The two next contributions can be merged into one term and by pulling $\pli{\te{R}}$ into the Poisson bracket, we get,
\begin{align}
\frac{i}{2}\, \lbrace \pli{\te{R}}\,\us{1}{}, \qf{h} + \im{E}{n} \rbrace \,\us{0}{\ast} \,\pli{\te{R}} &= \frac{1}{4}\, \left\lbrace  \lbrace  \pli{\te{R}}\,\us{0}{} , \qf{h} + \im{E}{n} \rbrace (\qf{h}-\im{E}{n})^{-1} (\Uf{\Hs{\te{p}}}\!\!-\pli{0}) , \qf{h} + \im{E}{n} \right\rbrace \,\us{0}{\ast} \,\pli{\te{R}}
\end{align}
Note that this expression involves the twofold application of the Poisson bracket with respect to $\KGv$ and $\KGm$ which makes the evaluation cumbersome. For completeness, we provide the result in terms of the connection entries,
{\footnotesize
\begin{align}
&\frac{i}{2}\,\pli{\te{R}}\,\lbrace \us{1}{}, \qf{h} + \im{E}{n} \rbrace\,\us{0}{\ast}\,\pli{\te{R}} = \frac{1}{4}\,\sum_{\delta} \!e_{\nD}^{\qpbzero} \scpr{e_{\nD}^{\qpbzero}}{\cdot} \nonumber \\
&~~  \!\!\cdot \left\lbrace  \sum_{m \neq \nD} \left\lbrace \left( (\!\mathcal{A}^{\scriptscriptstyle{\KGv}}\!)_{\nD}^{~m} (\!\mathcal{A}^{\scriptscriptstyle{\KGv}}\!)_{m}^{~\nD} \left(\!\frac{\partial (\im{E}{m}\!\!+\! \im{E}{n})}{\partial \KGm} \!\right)^{\!2} \!- (\!\mathcal{A}^{\scriptscriptstyle{\KGm}}\!)_{\nD}^{~m} (\!\mathcal{A}^{\scriptscriptstyle{\KGv}}\!)_{m}^{~\nD} \frac{\partial (\im{E}{m}\!\!+\! \im{E}{n})}{\partial \KGv} \frac{\partial (\im{E}{m}\!\!+\!\im{E}{n})}{\partial \KGm} \right) + \left( \KGv \leftrightarrow \KGm \right) \right\rbrace \frac{1}{\im{E}{m}\!\!- \!\im{E}{n}} \right. \nonumber \\
&~~~~ +  \sum_{m \neq \nD} \left\lbrace \left( \left( \frac{\partial (\!\mathcal{A}^{\scriptscriptstyle{\KGv}}\!)_{\nD}^{~m}}{\partial \KGv} (\!\mathcal{A}^{\scriptscriptstyle{\KGm}}\!)_{m}^{~\nD} - \frac{(\!\mathcal{A}^{\scriptscriptstyle{\KGv}}\!)_{\nD}^{~m}}{\partial \KGm} (\!\mathcal{A}^{\scriptscriptstyle{\KGv}}\!)_{m}^{~\nD} \right) \frac{\partial (\im{E}{m}\!+\!\im{E}{n})}{\partial \KGm} \right) +   \left( \KGv \leftrightarrow \KGm \right) \right\rbrace \nonumber \\
&~~~~ +\sum_{m \neq \nD} \left\lbrace  \left( (\!\mathcal{A}^{\scriptscriptstyle{\KGv}}\!)_{\nD}^{~m} (\!\mathcal{A}^{\scriptscriptstyle{\KGm}}\!)_{m}^{~\nD} \frac{\partial^2 (\im{E}{m}\!+\!\im{E}{n})}{\partial \KGv\,\partial \KGm} - (\!\mathcal{A}^{\scriptscriptstyle{\KGm}}\!)_{\nD}^{~m} (\!\mathcal{A}^{\scriptscriptstyle{\KGm}}\!)_{m}^{~\nD} \frac{\partial^2 (\im{E}{m}\!+\!\im{E}{n})}{\partial \KGv^2} \right)  +   \left( \KGv \leftrightarrow \KGm \right) \right\rbrace \nonumber\\
&~~~~ +\left. \sum_{j \neq \nD, \,m,\, l}\left\lbrace (\!\mathcal{A}^{\scriptscriptstyle{\KGv}}\!)_{\nD}^{~m} (\!\mathcal{A}^{\scriptscriptstyle{\KGm}}\!)_{m}^{~j} \left((\!\mathcal{A}^{\scriptscriptstyle{\KGv}}\!)_{j}^{~l} (\!\mathcal{A}^{\scriptscriptstyle{\KGm}}\!)_{l}^{~\nD}- (\!\mathcal{A}^{\scriptscriptstyle{\KGm}}\!)_{j}^{~l} (\!\mathcal{A}^{\scriptscriptstyle{\KGv}}\!)_{l}^{~\nD}  \right) \frac{\im{E}{m}\cdot (\im{E}{l}-\im{E}{n})}{\im{E}{j}- \im{E}{n}} \right\rbrace   \right\rbrace \label{eq:heff2 explicit}
\end{align}}

Here, $(\KGv \leftrightarrow \KGm)$ means that we replace every occurrence of $\KGv$ in the previous expression by $\KGm$, and vice versa.\\

However lengthy this expression might be, it includes only a few contributions which eventually enter at orders $\pp^0$ or lower after a careful examination of the terms. Therefore, we have a closer look at the connections and the energies and their respective orders in $\pp$. First realize that in the second line, we have an inverse factor of $(\im{E}{m}\!-\im{E}{n})$ with $m\neq \nD$ at the end. Since the difference between two eigenenergies of the perturbation system is always proportional to $\pp^2$ and higher orders, this causes a factor $\pp^{-2}$ in the second line. In the first factor of this line, we notice that whenever $(\!\mathcal{A}^{\scriptscriptstyle{\KGv}}\!)$ entails a shift with respect to the tensor modes, it enters with a factor $\pp^2$, because it is proportional to $\partial_{\KGv}(\pp^2\,\mT^2) \propto \pp^2$. As it turns out, this fact shifts all contributions involving the tensor modes into higher orders with respect to $\pp$, such that only the Mukhanov-Sasaki mode shifts will have an effect for the back reactions. On the other hand, the shifts produced by $(\!\mathcal{A}^{\scriptscriptstyle{\KGv}}\!)$ for the Mukhanov-Sasaki modes are proportional to $\partial_{\KGv}(\mMS^2) \propto \pp^1$. Regarding the entries of $(\!\mathcal{A}^{\scriptscriptstyle{\KGm}}\!)$, we recall that there are only vanishing contributions for tensor shifts, while for the Mukhanov-Sasaki modes, the connection is proportional to $\partial_{\KGm}(\mMS^2) \propto \pp^0$. We then come to the perturbation orders of the energy derivatives. The sum of the energies in the second line produce, according to the possible values of $m$, a sum of $2\,\im{E}{n}$ plus the respective energy shift. Since the energy shifts enter at least at order $\pp^2$, these contributions are not of further interest. Also the tensor and Mukhanov-Sasaki contributions in $2\,\im{E}{n}$ enter with $\pp^2$. Only the derivative with respect to $\KGm$ of the homogeneous part of the energy, $\irm{E}{hom}$, is of order $\pp^0$, while the derivative with respect to $\KGv$ enters with a factor $\pp^2$ as well. Combining these $\pp$-factors with the connections and the inverse energy contribution, we see that the only term which enters at zeroth order, is the one proportional to $(\!\mathcal{A}^{\scriptscriptstyle{\KGv}}\!)_{\nD}^{~m} (\!\mathcal{A}^{\scriptscriptstyle{\KGv}}\!)_{m}^{~\nD} (\partial_{\KGm} \irm{E}{hom})^2$. The explicit evaluation of the connections with respect to all possible wave vectors $\vec{m}$ for this term yields,
\begin{equation} \label{eq:heff2 relevant}
\sum_{\delta} \efzero{\nD} \,\langle \efzero{\nD} , \cdot \rangle \sum_{\vec{m}} \frac{1}{\left( \vec{m}^2 + \mMS^2 \right)^{5/2}} \left(2\,\nMS{m} +1\right)\, \left(- \frac{9}{2}\,\mPhi^4 \frac{\KGm^4}{\Gv^3 \Gm^2} \right).
\end{equation}
Note that this contribution finally enters with a factor $\pp^2$ from the overall perturbation order. Finally, we can perform the same careful analysis for all other terms in \eqref{eq:heff2 explicit} with the result that all other terms enter in \emph{higher} orders in $\pp$, and are thus, not of interest for the restriction at second perturbative order. \\

We come back to the remaining contributions to $\heff{,(2),\te{r}}{}$ in \eqref{eq:heff2 app}. After the examination of the first three terms, the fourth term is a trivial exercise: it involves the factor $\pli{\te{R}}\,\heff{,1}{}$ in the left-hand factor of the Poisson bracket, but we have already shown with \eqref{eq:heff1 vanishes} and \eqref{eq:piRheff1} that this is zero. \\

The last two contributions involve the terms of the Moyal product at second order, which give,
{\footnotesize
\begin{align}
\pli{\te{R}}\left((\us{0}{} \spr \qf{h})_2 - (\im{E}{n} \spr \us{0}{})_2 \right)\,\us{0}{\ast}\,\pli{\te{R}} =  - \frac{1}{8}\, \frac{\partial^2 \us{0}{}}{\partial \KGv^2} \,\frac{\partial^2 (\qf{h}-\!\im{E}{n})}{\partial \KGm^2} + \frac{1}{4}\, \frac{\partial^2 \us{0}{}}{\partial \KGv\, \partial \KGm} \,\frac{\partial^2 (\qf{h}-\!\im{E}{n})}{\partial \KGm\,\partial \KGv} - \frac{1}{8}\, \frac{\partial^2 \us{0}{}}{\partial \KGm^2} \,\frac{\partial^2 (\qf{h}-\!\im{E}{n})}{\partial \KGv^2}.
\end{align}}

The explicit evaluation in terms of the connections gives,
\begin{align}
& \pli{\te{R}}\left((\us{0}{} \spr \qf{h})_2 - (\im{E}{n} \spr \us{0}{})_2 \right)\,\us{0}{\ast}\,\pli{\te{R}} = \frac{1}{4} \sum_{\delta} \efzero{\nD} \scpr{\efzero{\nD}}{\cdot} \nonumber \\
&~~~~  \cdot \left\lbrace \sum_{m} \left\lbrace \left( \frac{\partial (\im{E}{m}\!\!- \!\im{E}{n})}{\partial \KGm} \left( (\!\mathcal{A}^{\scriptscriptstyle{\KGv}}\!)_{\nD}^{~m}\frac{\partial (\!\mathcal{A}^{\scriptscriptstyle{\KGm}}\!)_{m}^{~\nD}}{\partial \KGv} - (\!\mathcal{A}^{\scriptscriptstyle{\KGm}}\!)_{\nD}^{~m}\frac{\partial (\!\mathcal{A}^{\scriptscriptstyle{\KGv}}\!)_{m}^{~\nD}}{\partial \KGv} \right) \right) + \left(\KGv-\KGm\right) \right\rbrace \right. \nonumber \\
&~~~~~~ + \frac{1}{2} \sum_m \left\lbrace (\im{E}{m}\!\!-\!\im{E}{n}) \left(\frac{\partial (\!\mathcal{A}^{\scriptscriptstyle{\KGv}}\!)_{\nD}^{~m}}{\partial \KGm} \frac{\partial (\!\mathcal{A}^{\scriptscriptstyle{\KGm}}\!)_{m}^{~\nD}}{\partial \KGv} - \frac{\partial (\!\mathcal{A}^{\scriptscriptstyle{\KGv}}\!)_{\nD}^{~m}}{\partial \KGv} \frac{\partial (\!\mathcal{A}^{\scriptscriptstyle{\KGm}}\!)_{m}^{~\nD}}{\partial \KGm} \right) + (\KGv \leftrightarrow \KGm)\right\rbrace \nonumber \\
&~~~~~~ + \sum_{m,k,j} \left\lbrace \left( \im{E}{k} - \im{E}{j} - \im{E}{m} - \im{E}{n}\right) (\!\mathcal{A}^{\scriptscriptstyle{\KGv}}\!)_{\nD}^{~m} (\!\mathcal{A}^{\scriptscriptstyle{\KGv}}\!)_{m}^{~\nD} (\!\mathcal{A}^{\scriptscriptstyle{\KGm}}\!)_{k}^{~j} (\!\mathcal{A}^{\scriptscriptstyle{\KGm}}\!)_{j}^{~\nD}\right\rbrace \nonumber \\
&~~~~~~ + \left. \sum_{m,k,j} \left\lbrace (\!\mathcal{A}^{\scriptscriptstyle{\KGm}}\!)_{\nD}^{~m} (\!\mathcal{A}^{\scriptscriptstyle{\KGm}}\!)_{m}^{~k} \left( (\!\mathcal{A}^{\scriptscriptstyle{\KGm}}\!)_{k}^{~j} (\!\mathcal{A}^{\scriptscriptstyle{\KGv}}\!)_{j}^{~\nD} (\im{E}{n}\!-\!\im{E}{j}) +  (\!\mathcal{A}^{\scriptscriptstyle{\KGv}}\!)_{k}^{~j} (\!\mathcal{A}^{\scriptscriptstyle{\KGm}}\!)_{j}^{~\nD} (\im{E}{k}\!-\!\im{E}{j}) \right) \right\rbrace \right\rbrace.
\end{align}
We identify some of the terms that already appeared in the second contribution of $\heff{,2,\te{r}}{}$. The same analysis with respect to the $\pp$-factors reveals that none of the terms enters at zeroth order in $\pp$. \\

The last term of $\heff{,2,\te{r}}{}$ corresponds to an effective Hamiltonian of first order for the gravitational Moyal product. The same line of arguments as for $\heff{,1,\te{r}}{}$ in \eqref{eq:heff1 vanishes} applies, and consequently this contribution vanishes. \\

Hence, the only relevant contribution of $\heff{,2,\te{r}}{}$ for our perturbative treatment remains the term in \eqref{eq:heff2 relevant}. \\

\end{appendix}

\end{document}